\documentclass[aps,prd,nofootinbib,superscriptaddress]{revtex4}

\usepackage{amsfonts,amssymb,amsthm,bbm,amsmath}
\usepackage{color,psfrag}
\usepackage[dvips]{graphicx}

\newcommand{\C}{{\mathbb C}}
\newcommand{\N}{{\mathbb N}}
\newcommand{\R}{{\mathbb R}}
\newcommand{\Z}{{\mathbb Z}}

\newcommand{\cB}{{\mathcal B}}

\newcommand{\cV}{{\mathcal V}}

\newcommand{\cH}{{\mathcal H}}

\newcommand{\cD}{{\mathcal D}}

\newcommand{\SU}{\mathrm{SU}}

\newcommand{\be}{\begin{equation}}
\newcommand{\ee}{\end{equation}}
\newcommand{\beq}{\begin{eqnarray}}
\newcommand{\eeq}{\end{eqnarray}}
\newcommand{\bea}{\begin{eqnarray}}
\newcommand{\eea}{\end{eqnarray}}
\newcommand{\nn}{\nonumber}

\newcommand{\mat} [2] {\left ( \begin{array}{#1}#2\end{array} \right ) }

\newcommand{\su}{\mathfrak{su}}

\renewcommand{\sl}{\mathfrak{sl}}

\newcommand{\la}{\langle}
\newcommand{\ra}{\rangle}

\newcommand{\f}{\frac}

\def\nn{\nonumber}
\def\pp{\partial}
\def\arr{\rightarrow}

\def\vphi{\varphi}
\def\eps{\epsilon}
\def\dag{^\dagger}

\newcommand{\id}{\mathbb{I}}

\def\bz{\bar{z}}

\def\balpha{\bar{\alpha}}
\def\bbeta{\bar{\beta}}

\def\vu{\vec{u}}

\def\vJ{\vec{J}}

\def\vsigma{\vec{\sigma}}

\def\tn{{\tilde{n}}}
\def\ttau{{\tilde{\tau}}}

\begin{document}

\title{Group theoretical Quantization of Isotropic Loop Cosmology}

\author{{\bf Etera R. Livine}}\email{etera.livine@ens-lyon.fr}
\affiliation{Laboratoire de Physique, ENS Lyon, CNRS-UMR 5672, 46 All\'ee d'Italie, Lyon 69007, France}

\author{{\bf Mercedes Mart\'\i n-Benito}}\email{mercedes@aei.mpg.de}
\affiliation{MPI f. Gravitational Physics, Albert Einstein Institute, Am M\"uhlenberg 1, D-14476 Potsdam, Germany}

\date{\today}

\begin{abstract}
We achieve a group theoretical quantization of the flat Friedmann-Robertson-Walker model coupled to a massless scalar field adopting the improved dynamics of loop quantum cosmology. Deparemeterizing the system using the scalar field as internal time,
we first identify a complete set of phase space observables whose Poisson algebra is isomorphic to the $\su(1,1)$ Lie algebra.
It is generated by the volume observable and the Hamiltonian.
These observables describe faithfully the regularized phase space underlying the loop quantization: they account for the polymerization of the variable conjugate to the volume and for the existence of a kinematical non-vanishing minimum volume.
Since the Hamiltonian is an element in the $\su(1,1)$ Lie algebra, the dynamics is now implemented as $\SU(1,1)$ transformations.
At the quantum level, the system is quantized as a time-like irreducible representation of the group $\SU(1,1)$. These representations are labeled by a half-integer spin, which gives the minimal volume. They provide superselection sectors without quantization anomalies and no factor ordering ambiguity arises when representing the Hamiltonian.
We then explicitly construct $\SU(1,1)$ coherent states to study the quantum evolution. They not only provide semiclassical states but truly dynamical coherent states. Their use further clarifies the nature of the bounce that resolves the big bang singularity.

\end{abstract}

\maketitle
\tableofcontents

\section*{Introduction}

In the last decade loop quantum cosmology (LQC) has been established as a promising model of quantum cosmology in its attempt to address some of the fundamental issues of standard cosmology, such as the avoidance of the initial singularity, origin of inflation, etc. For recent reviews see \cite{Boj,AsS,BCM}.
The paradigmatic model in LQC is the flat Friedmann-Robertson-Walker (FRW) model coupled to a homogeneous massless scalar field, the simplest cosmological model with non-trivial dynamics. Actually, owing to the isotropy and the homogeneity, the model is constrained by a single global Hamiltonian constraint. Furthermore, the matter term corresponding to the massless scalar allows us to deparameterize the system, regarding the scalar as the internal time, and easily solve it. Solutions undergo a singularity at vanishing volume of the universe.

In his pioneering works \cite{boj1a,boj1b,boj1c,boj1d,boj2}, Bojowald proposed
to adapt the quantization techniques of loop quantum gravity \cite{lqg1,lqg2,lqg3} to construct a singularity-free quantization of this simplest model. Then, the mathematical structure of LQC was rigorously established \cite{abl,Vel} and the quantization of the model was completed \cite{aps1,aps2,aps3}. The improved dynamics introduced in \cite{aps3} showed that, as desired, the big bang singularity is resolved being replaced by a quantum bounce: choosing as physical observable the volume at a given value of the internal time, then while the time varies, the expectation value of the volume in physical states features a contraction epoch, till it bounces to start expanding. In the moment of the bounce the matter density reaches a finite maximum value that is of Planck order.

Though the quantization of the model was successfully completed in \cite{aps3}, it has been further investigated. Essentially, playing around with the factor ordering ambiguity present when symmetrizing the Hamiltonian constraint operator, different orderings, with different advantages with respect to the original ordering of \cite{aps3}, have been proposed (see e.g. \cite{acs,mmo,mop}), all of them with the same asymptotic behavior but different at small scales.
These analysis show that the bounce featured by the quantum evolution holds for all choices of factor ordering and is universal: it happens for all the physical states.

Moreover, following the cosmic recall scenario originally proposed in \cite{cs} and further developed as a generic feature of loop quantum cosmology in \cite{kp-posL}, it has been shown that under certain conditions the bounce preserves semi-classicality : the expectation value of the volume in states that are semiclassical at late times follows as time varies a well defined trajectory with bounded relative fluctuations.
As a result it is possible to derive an effective classical dynamics generating those trajectories (see e.g. \cite{Tav}).
Actually, this effective dynamics can be understood as a consequence of a process of phase space regularization sometimes called ``polymerization'': given the basic variable describing the geometry, the volume in the case of the improved dynamics of LQC, denoted by $v$, its canonically conjugate variable $b$ is regularized by the expression $\sin(\lambda b)/\lambda$ with $\lambda$ a fixed real parameter with dimension of a length (usually set to the Planck scale). As a result, the phase space is described by $v$ and by the exponentiated observables $e^{\pm i\lambda b}$, instead of $b$ itself. In this regularization lies the bounce mechanism solving the singularity.\footnote{The regularized algebra generated by $v$ and  $e^{\pm i\lambda b}$ is an adaptation  to this homogeneous situation of the regularized holonomy-flux algebra employed in loop quantum gravity.}
\\

In this work we look again at the flat FRW model coupled to a massless scalar, within the improved dynamics of LQC, improving further the quantization. Now we propose a different and more natural approach, following group theoretical techniques. Indeed, it is easy to realize that the Poisson algebra of a basic set of observables describing the phase space in LQC is isomorphic to the $\su(1,1)$ algebra.
This property was already pointed out in \cite{boj3}, and more recently in \cite{2vertex} in the context of the dipole cosmology model derived from loop quantum gravity and spinfoam models \cite{francesca1, francesca2, sfcosmo}. In \cite{2vertex}, although the $\su(1,1)$ structure was discussed to have a more fundamental role, it was merely used to deduce the spectrum of the Hamiltonian. On the other hand, in \cite{boj3}, the $\su(1,1)$ group structure was used in a deeper way to derive the quantum cosmological evolution.
However in that work the advantages of having a consistently quantizable algebra were not fully exploited, since no use of the group theoretical quantization was employed. Rather, the model was analyzed from an algebraic point of view: the quantum evolution was not derived from a process of quantizing the observables on a Hilbert space, but was described as a set of coupled classical equations of motion for the expectation values, fluctuations and correlations of the observables.
This allowed to study the cosmological evolution of the fluctuations, but did not provide an explicit analysis of the Hilbert space and quantum states of geometry.

Instead, we here take full advantage of the $\su (1,1)$ algebra structure of the model. We will perform a transparent quantization, simply by representing the set of phase space observables as self-adjoint operators associated with the generators of the algebra. In this way, the different superselection sectors that our quantization features will correspond to the irreducible representations of the group $\SU(1,1)$ of the discrete principal series. In comparison with the standard LQC procedure of e.g. \cite{aps3,acs,mmo, mop}, several advantageous novelties follow:
\begin{itemize}
\item In the usual LQC approach the Hamiltonian operator suffers from factor ordering ambiguities. Rather, in our description, the Hamiltonian is an element of the $\su(1,1)$ algebra and thus it is represented without ambiguity. Moreover, the evolution is simply generated by $\SU(1,1)$ transformations.

\item In LQC the volume variable $v$ lies in the real line. It is defined from the triad variable through a canonical transformation and its sign reflects the orientation of the triad. Strictly speaking, its absolute value $|v|$ gives the volume (up to a numerical factor). Then one would wish to restrict to positive values of $v$ in a way consistent with the dynamics, in order to avoid unphysical cross-overs and interferences between positive and negative orientation states. Previous works attained the decoupling of positive  values of $v$ from negative ones either appealing to parity symmetry \cite{aps3,acs}, or proposing a suitable factor ordering for the Hamiltonian constraint \cite{mmo,mop}. In our approach this is no longer an issue: positive (negative) values of $v$ correspond to the  positive (negative) discrete principal series of $\SU(1,1)$. Therefore positive and negative values of $v$ are decoupled beforehand, each sector providing an irreducible representation.

\item In LQC the kinematical volume $v$ is discrete, due to the nature of the loop quantization. Moreover it is superselected in decoupled sectors. In each sector the admissible values of $v$ form a lattice of equidistant points \cite{aps3}. Therefore,
once $v$ is restricted to be positive, it features a minimum non-vanishing value characteristic of the corresponding superselection sector.
The studies of the effective dynamics proposed so far in the literature (see e.g. \cite{aps3,Tav}) ignore this fact, and they only take into account the polymerization of the conjugate variable $b$, assuming $v\geq 0$. In comparison, our approach can be consistently generalized to account, not only for the regularization of the variable $b$, but also for the regularization of the volume, such that at the classical level $v\geq v_m>0$. Namely, we can really describe the regularized phase space underlying LQC accounting for the existence of a minimal volume directly at the classical level.

\item The kinematical minimum volume labels different superselection sectors in the quantum theory. In the usual LQC approach this label takes values in a continuous finite interval. In contrast, in our approach, the kinematical minimum volume is discrete, since its value is the spin $j=(1+\mathbb{N})/2$ that labels the chosen time-like irreducible representation of the group $\SU(1,1)$. Then, unlike in usual LQC, the direct sum of our superselection sectors is still a separable Hilbert space.

\item So far, in the previous quantization schemes of the model within LQC, semi-classical states were provided \cite{aps3,mop,CoM}, but not truly coherent, since those states changed shape under evolution and did not saturate the uncertainty relations. In our case, $\su(1,1)$ coherent states will naturally provide explicit and exact dynamical coherent states. The analysis of the expectation values and fluctuations of physical observables in these states confirms once again the universality of the quantum bounce and that fluctuations remain bounded. Let us note that in \cite{boj3} coherent states were also discussed. Although not explicitly constructed, the evolution of their fluctuations and correlations was derived. We will compare our results with those of \cite{boj3}.

\item The group theoretical perspective provides a rigorous setting when analyzing possible generalizations of the FRW model. If other terms such as curvature or cosmological constant admit a description in terms of the elements of the algebra, then they will also admit an anomaly free quantization.
\end{itemize}

The structure of the paper is as follows. In section \ref{Sec1} we review the classical flat FRW model in the presence of a massless scalar and then the corresponding effective model derived from LQC, regularized by taking into account the polymerization of $b$. In section \ref{Sec2} we describe the $\SU(1,1)$ group structure of this effective model and explicitly show that the evolution is given by $\SU(1,1)$ transformations. In section \ref{Sec3} we modify the previous description in order to take into account also the regularization of the volume and, in this way, consider the fully regularized classical model underlying LQC. This fully regularized model is then quantized in section \ref{Sec4} by considering the time-like representations of $\SU(1,1)$. We explicitly construct dynamical coherent states and use them to analyze the quantum evolution. We also compare our approach with previous quantizations of the model. In section \ref{Sec5} we generalize our analysis by considering a generic $\su(1,1)$ Hamiltonian, in order to study whether curvature or cosmological constant can be implemented simply in our framework. Finally we conclude summarizing the main results of this work.

We detail in two appendices the group theoretical tools employed in the paper. We review the Schwinger representation of the classical $\su(1,1)$ algebra in appendix \ref{AppA}. Then we construct the time-like irreducible representations of $\SU(1,1)$ and provide coherent states together with their properties  in appendix \ref{AppB}. Finally appendix \ref{AppC} reviews the classical description of the FRW model with curvature or cosmological constant.


\section{Classical and Regularized FRW Models}
\label{Sec1}

In this section we will briefly review the Hamiltonian formulation of the flat FRW model in the presence of a massless scalar.
We will start by the classical model within general relativity. Then, we will review how this classical dynamics is modified when considering the regularization employed in loop quantum cosmology\footnote{Along this paper we work with units $\hbar=c=1$.}.

\subsection{Standard Flat FRW Cosmology}
The (standard) flat FRW model represents isotropic and homogeneous solutions of the Einstein equations with flat spatial sections.
Since these spatial sections are non-compact, and the variables that describe the model are spatially homogeneous, several integrals that appear in the Hamiltonian framework, such as the symplectic structure or the spatial average of the Hamiltonian constraint, diverge. To avoid these divergences, one usually restricts the analysis to a finite cell $\mathcal V$. Owing to the homogeneity, the study of this cell reproduces what happens in the whole universe.

Thanks to the homogeneity and the isotropy of the model, the geometry sector of the phase space can be described by a single pair of canonical variables. Usually one employs the scale factor $a$ and its canonically conjugate momentum $\pi_a$, such that $\{a,\pi_a\}=1$. On the other hand, let us denote by $\phi$ the massless scalar, and by $p_\phi$ its momentum, such that $\{\phi,p_\phi\}=1$. Owing to the homogeneity, this phase space is only constrained by the scalar or Hamiltonian constraint, which reads
\begin{align}
C=-\frac{2\pi G}{3}\frac{\pi_a^2}{a}+\frac{p_\phi^2}{2a^3}=0\,.
\end{align}

In LQC, following loop quantum gravity, the phase space of the model was originally described by a real coefficient $c$ parameterizing the Ashtekar-Barbero connection (which encodes the extrinsic curvature), and a real coefficient $p$ parameterizing the densitized triad (which measures the area), defined such that $\{c,p\}=8\pi G\gamma/3$, being $\gamma$ the Immirzi parameter \cite{abl}. This set of variables is related with the previous one by the canonical transformation
\begin{align}
a=\sqrt{|p|}\,,\qquad \pi_a=-\frac{3}{4\pi G \gamma}\text{sign}(p)\sqrt{|p|}c\,,
\end{align}
so that the scalar constraint in these variables becomes
\begin{align}
C=\frac{1}{16\pi G}\left[-\frac{6}{\gamma^2}c^2\sqrt{|p|}+8\pi G\frac{p_\phi^2}{|p|^{3/2}}\right]=0\,.
\end{align}

\begin{figure}[h]
\begin{center}
\includegraphics[height=35mm]{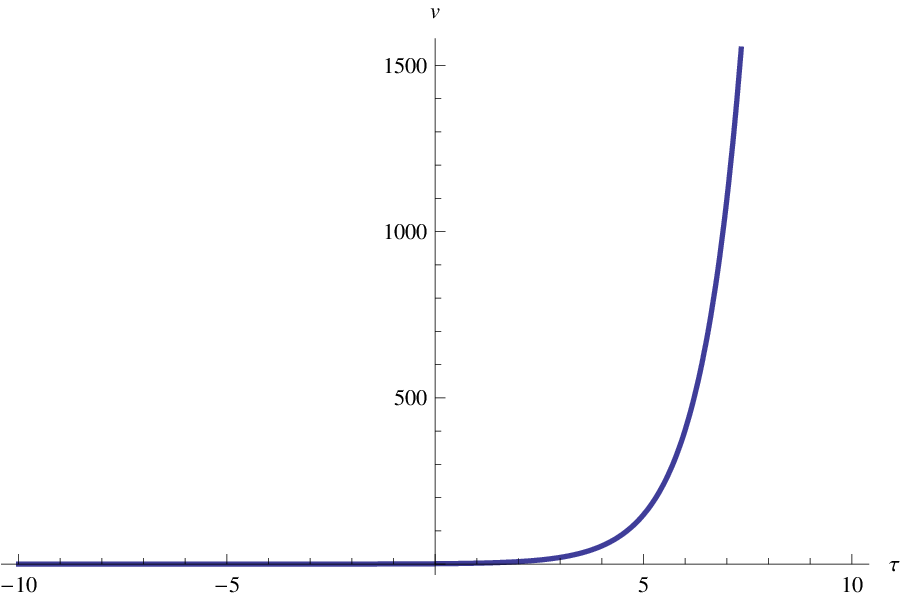}
\hspace{10mm}
\includegraphics[height=35mm]{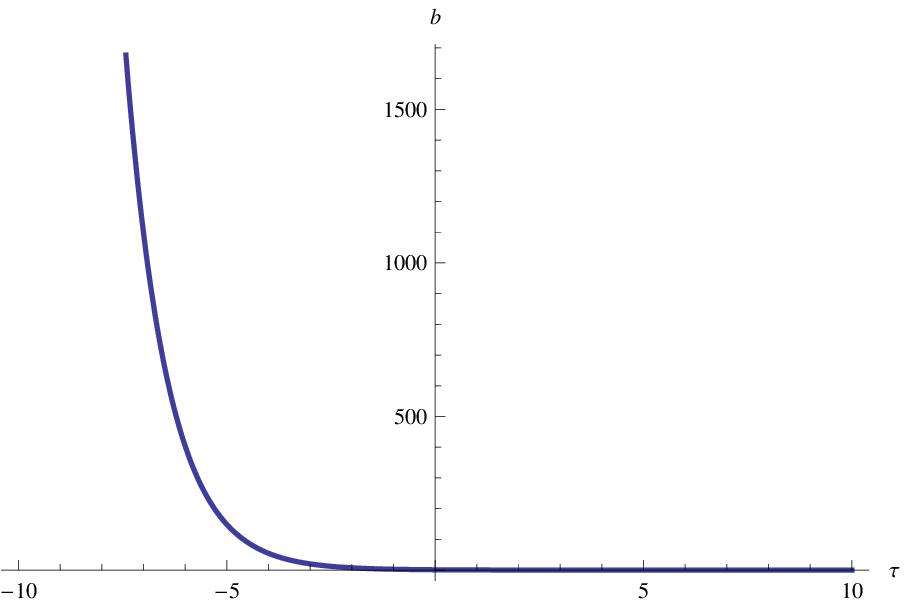}
\caption{Plots of the volume $v$ (on the left) and of its conjugate momentum $b$ (on the right) evolving as functions of the internal time $\tau$, for $\tau_o=0$, for the negative branch: the universe starts with a big bang at $\tau\arr-\infty$ to expand to infinite volume  $\tau\arr+\infty$.
\label{frw}}
\end{center}
\end{figure}

The improved dynamics scheme \cite{aps3} proved later that it is better to describe the geometry in terms of the volume instead of the area.\footnote{In this way, the polymeric representation of the resulting algebra leads to a quantum evolution in agreement with general relativity at semiclassical scales and introducing important quantum effects only at Planck scales. In turn, the ``old dynamics'', in which the polymeric representation is carried out using the variables $p$ and $c$, led to the possibility of having important quantum effects at classical scales \cite{aps1,aps2}.} Then, one introduces a variable $v$ measuring the volume $V$ of the cell under study, $V=4\pi G |v|$, and its canonically conjugate variable $b$, such that $\{b,v\}=1$.\footnote{Note that, according with these definitions, $v$ and $b^{-1}$ have dimensions of length. Then $V=\propto Gv$ has correctly the dimension of a volume.} The relation between these variables and the previous ones is
\begin{align}
p&=\text{sign}(v)\left(4\pi G |v|\right)^{2/3}\,, \qquad c=\gamma \left(4\pi G |v|\right)^{1/3} b\,.\\
a&=(4\pi G|v|)^{1/3}\,,\qquad\qquad\;\; \pi_a=-3\text{sign}(v)\left(\frac{v^2}{4\pi G}\right)^{1/3}b\,.
\end{align}
Therefore, the Hamiltonian constraint is given by
\begin{align}
 C=-\frac{3}{2} b^2v+\frac{p_\phi^2}{8\pi Gv}=0\,.
\end{align}
It is obvious to realize that $p_\phi$ is a constant of motion, since it Poisson commutes with the Hamiltonian constraint. Then we can deparameterize the system by regarding $\phi$ as an internal time and $p_\phi$ as the (physical) Hamiltonian which generates evolution in the time $\phi$. In view of the constraint we have
\begin{align}
p_\phi=\pm\sqrt{12\pi G} bv\equiv H_\pm\,.
\end{align}
We obtain two branches of solutions. Let us first consider the negative branch. We define the time parameter $\tau\equiv \sqrt{12\pi G}\phi$ for convenience, then the equations of motion have a very simple form,
\begin{align}
\partial_\tau v=\frac{1}{\sqrt{12\pi G}}\{v,H_-\}=v(\tau)\,,\qquad \partial_\tau b=\frac{1}{\sqrt{12\pi G}}\{b,H_-\}=-b(\tau)\,,
\end{align}
with solution
\begin{align}
v(\tau)=v(\tau_o)e^{\tau-\tau_o}\,,\qquad b(\tau)=b(\tau_o)e^{-(\tau-\tau_o)}\,.
\end{align}
As shown on fig. \ref{frw}, the solutions correspond to a universe expanding from a vanishing volume at $\tau\rightarrow -\infty$ to an infinity volume at $\tau\rightarrow \infty$.
In consequence, the matter density $\rho_\phi=p_\phi^2/2V^2\propto p_\phi^2/v^2$ diverges at initial time $\tau\rightarrow -\infty$, which corresponds to the initial big bang singularity. Then the volume grows as the scalar field grows too.
The negative branch is the time reversal of the positive branch. Thus it consists in solutions contracting from infinite volume to vanishing volume, where a big crunch singularity is formed.

It is convenient for some purposes to switch back to proper time $t$ for which the evolution is given by taking the constraint $C$ as the Hamiltonian.
Then the equation of motion of $\phi$ in terms of the proper time $t$ is given by $d{\phi}/dt=\{\phi,C\}$, so that the relation between the proper time and the internal time is
\be\label{propert}
dt=\f{4\pi G}{p_\phi} v d\phi = \f{1}{p_\phi}\sqrt{\f {4\pi G}3}\,v\,d\tau.
\ee
Considering the negative branch, this is easily integrated, setting $\tau_0=0$ for simplicity's sake:
\be
t
\,=\,
\f{v_0}{p_\phi}\sqrt{\f {4\pi G}3}\,e^\tau
\,=\,
\f{1}{p_\phi}\sqrt{\f {4\pi G}3}\,v\,,
\ee
so that the internal time evolution $\tau\in\R$ is mapped onto positive proper time $t\in]0,+\infty[$ and the initial proper time $t=0$ corresponds to $\tau\arr-\infty$ and vanishing volume, that is to the big bang singularity.

Interestingly, the expansion rate in internal time is constant, since $\pp_\tau v=v$. But converting this back to proper time, we recover the standard Hubble expansion rate given by:
\be
\f{\pp_t a}{a}=\f13\f{\pp_t v}{v}=\f13\f{\pp_\tau v}{v}\,\f{d\tau}{dt}
\quad=\pm
\f{p_\phi}{\sqrt{12\pi G}}\,\f1 v
\quad=\,
b\,.
\ee
In particular, this allows to recover the standard Friedmann equation:
\be
\left(\f{\pp_t a}{a}\right)^2
=\f{8\pi G}{3}\,\rho\,.
\ee

\subsection{Effective FRW Cosmology from LQC}


LQC successes in solving the cosmological singularity essentially owing to a process of regularization. Indeed, the basic observables describing the geometry are chosen to be the variable $v$ and the exponentials $e^{i\lambda b}$, with fixed length scale $\lambda$, instead of $b$ itself.\footnote{Unlike in a standard Schr\"{o}dinger quantization, in LQC the Hilbert space is not the space of smooth functions of the configuration variable $b$, square integrable with respect to the Lebesgue measure. Rather, the Hilbert space of LQC is the Bohr compactification of the real line  \cite{abl,Vel}. A basis of this space is provided by the almost periodic functions of $b$, whose elements are linear combinations of exponentials $e^{i\lambda b}$ with $\lambda\in\mathbb{R}$. Hence, they describe the configuration space. The regularization of the curvature tensor later requires to fix the value of $\lambda$ to a constant of Planck order \cite{aps3}.} Then $b$ is regularized by the expression \cite{aps3}
\begin{align}
\frac{\sin(\lambda b)}{\lambda}=\frac{e^{i\lambda b}-e^{-i\lambda b}}{2i\lambda}\,,
\end{align}
and thus the regularized Hamiltonian is given by
\begin{align}
H_\text{eff}^{\pm}\equiv p_\phi=\pm\sqrt{12\pi G}\frac{\sin(\lambda b)}{\lambda}v\,.
\end{align}
Considering the negative branch, the equations of motion now are
\begin{align}
\partial_\tau v=\cos(\lambda b)v\,,\qquad \partial_\tau b=-\frac{\sin(\lambda b)}{\lambda}\,,
\end{align}
with solution
\begin{align}\label{sol}
v(\tau)=v_o\cosh(\tau-\tau_o)\,,\qquad b(\tau)=\frac{1}{\lambda}\arccos[\tanh(\tau-\tau_o)]\,,
\end{align}
where $\tau_o$ and $v_o$ are constants of integration.

\begin{figure}[h]
\begin{center}
\includegraphics[height=35mm]{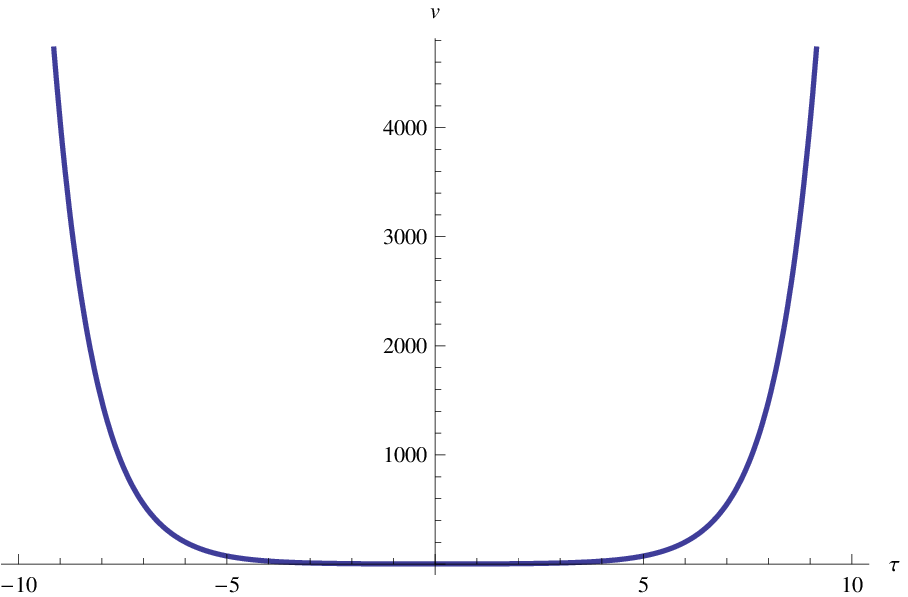}
\hspace{10mm}
\includegraphics[height=35mm]{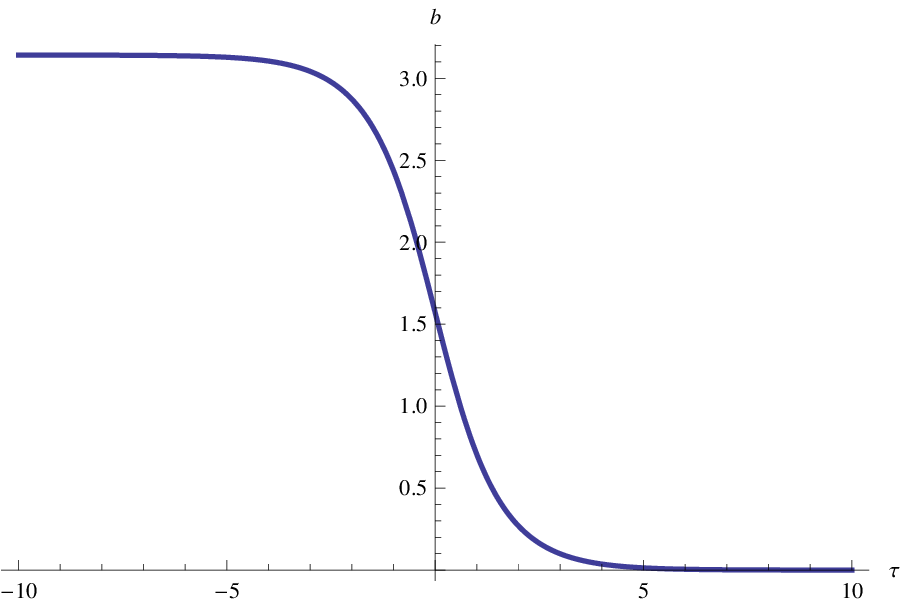}
\caption{Plots of the volume $v$ (on the left) and of its conjugate momentum $b$ (on the right) evolving as functions of the internal time $\tau$, for $\tau_o=0$, for the effective loop quantum dynamics of flat FRW cosmology: the universe starts with an infinite volume at $\tau\arr-\infty$, contracts and bounces to expand again to infinite volume at $\tau\arr+\infty$.
\label{frw_eff}}
\end{center}
\end{figure}

Note that these solutions are invariant under time reversal, and therefore in this case positive and negative branches merge in a unique branch of solutions.
As we can see on fig. \ref{frw_eff}, these solutions correspond to a universe that contracts from infinite volume at $\tau\rightarrow -\infty$ till it reaches a minimum volume $v(\tau_o)=\lambda p_\phi/\sqrt{12\pi G}$, and then  starts expanding till infinite volume at $\tau\rightarrow +\infty$.
Therefore the universe suffers a bounce at $\tau=\tau_o$. The matter density,
\begin{align}
\rho_\phi(\tau)=\frac{p_\phi^2}{2[4\pi G v(\tau)]^2}=\frac{3}{8\pi G \lambda^2}\frac{1}{\cosh^2(\tau-\tau_o)}\,,
\end{align}
reaches a non-divergent maximum at the bounce, given by
\begin{align}
\rho_\phi(\tau_o)=\frac{p_\phi^2}{2[4\pi G v_o]^2}=\frac{3}{8\pi G \lambda^2}\quad\equiv\rho_c\,.
\end{align}
This maximal density is universal (independent of the momentum of the field) and of Planck order, inasmuch as $\lambda$ is of Planck order as well. The usual value employed in the literature is $\lambda=\sqrt{\Delta}\gamma$ being $\Delta:=4\sqrt{3}\pi\gamma l_\text{Pl}^2$ and $l_\text{Pl}^2$ the Planck length (see e.g. \cite{AsS} for the explanation of how the value of $\lambda$ is chosen).
In conclusion, the singularities present in the standard model are resolved here by a bounce mechanism.
Note that in the low extrinsic curvature regime,  $b(\tau)\rightarrow 0$, approached in the limits  $\tau\rightarrow \pm\infty$, we have $v(\tau)\rightarrow v_o e^{\pm\tau}$, so that the solutions tend respectively to the expanding and contracting solutions of the standard model reviewed in the previous subsection, and thus this effective dynamics is in agreement with general relativity in the semi-classical regime, far away from the bounce.

Similarly to before for the classical FRW universe, we can easily switch back to proper time:
\be
dt
\,=\,
\f{1}{p_\phi}\sqrt{\f {4\pi G}3}\,v\,d\tau
\quad\Rightarrow\quad
t=\f{v_o}{p_\phi}\sqrt{\f {4\pi G}3}\,\sinh \tau\,,
\ee
where we have set $\tau_o=0$ for simplicity's sake.
Now real internal time $\tau\in\R$ maps onto real proper time $t\in\R$ and we do not have a singularity anymore at $t=0$ but simply the bounce.
Finally this allows us to compute the modification of the Friedmann equation:
\be
\f{\pp_t a}{a}=\f13\f{\pp_\tau v}{v}\,\f{d\tau}{dt}
=\f{p_\phi}{\sqrt{12\pi G}}\,\f{\cos\lambda b} v
\quad\Rightarrow\quad
\left(\f{\pp_t a}{a}\right)^2=\f{8\pi G}{3}\,\rho\,\left(1-\f{\rho}{\rho_c}\right)\,,
\ee
where $\rho_c$ is the maximal/critical density defined above. The new factor on the right hand side is the leading order modification of the Friedmann equations in loop quantum cosmology.

\medskip

In the following, to simplify the notation, we will absorb the factor $\lambda$ by means of the canonical transformation
\begin{align}
 v\rightarrow v'=\frac{v}{\lambda}\,,\qquad b\rightarrow b'=\lambda b
\end{align}
and redefine the dimensionless variables $v\equiv v'$ and $b\equiv b'$.

\section{Group Structure of Effective FRW Cosmology}
\label{Sec2}
\subsection{The $\su(1,1)$ Algebra Governing the Dynamics}

As we have seen in the previous section, in LQC the physical phase space (after deparameterization) is described by the basic variables $v$ and $e^{\pm i b}$.
Combining the above basic variables, let us consider the set of observables
\begin{align}
 J_z=v, \qquad K_+=v e^{ib},\qquad K_-=v e^{-ib},
\end{align}
or alternatively the set of real observables
\begin{align}
 J_z= v\,,\qquad K_x=\frac{1}{2}(K_++K_-)= v \cos b\,, \qquad  K_y=\frac{1}{2i}(K_+-K_-)=v\sin b\,.
\end{align}
Using $\{b,v\}=1$, it is straightforward to realize that the Poisson algebra of these observables is a  $\sl_2\sim\su(1,1)$ Lie algebra:
\be
\{J_z,K_\pm\}=\,\mp i K_\pm,\qquad
\{K_+,K_-\}=\,2iJ_z\,,
\ee
\be
\{J_z,K_x\}=K_y,
\qquad
\{J_z,K_y\}=-K_x,
\qquad
\{K_x,K_y\}=-J_z\,.
\ee

The above  isomorphism between the Poisson algebra of observables with the  $\su(1,1)$ algebra induces an isomorphism between the group of canonical transformations on the phase space and the group of $SU(1,1)$ transformations.
Let us check that indeed the $SU(1,1)$ transformations can be seen as canonical transformations on phase space.
Given the $2\times 2$ matrix
\be
M\,\equiv\,
\mat{cc}{J_z & K_- \\ K_+ & J_z}\,=\left( \begin{array}{cc}
v & ve^{-ib} \\
ve^{ib} & v\end{array} \right),
\ee
whose determinant is the Casimir of the $\su(1,1)$ algebra
\be
C\,\equiv\, J_z^2-K_x^2-K_y^2\,=\,J_z^2-K_+K_- \,,
\ee
the transformations generated by a generic $SU(1,1)$ element $U$ read $M\rightarrow\tilde M=U M U^\dagger$, since $C=\det \tilde M=\det M$. See also Appendix \ref{AppA}, where we explicitly show that $M$ lives in the adjoint representation.
Parameterizing $U$ as
\be
U=\mat{cc}{\alpha & \beta \\ \bar{\beta} & \bar{\alpha}},
\qquad\textrm{with}\quad
|\alpha|^2-|\beta|^2=1\,,
\ee
it is easy to check that the above transformation induces the following transformation on the phase space variables\footnotemark~:
\begin{align}
v\,\rightarrow \,&\tilde v=v|\alpha+\beta e^{ib}|^2\,,\\
e^{ib}\,\rightarrow \,&
e^{i\tilde b}=\frac{\bar \alpha e^{ib}+\bar{\beta}}{\alpha+\beta  e^{i b}}\,,
\end{align}
which is canonical since $\{\tilde b,\tilde v\}=1$, as we wanted to prove.
\footnotetext{Let us notice that the transformation $e^{ib}\arr e^{i\tilde b}$ is a M\"obius transformation, which is a conformal transformation on the Riemann sphere and seems related to the Witt algebra generated by the observables $v e^{inb}$ for $n\in\Z$ generalizing our $\su(1,1)$ observables.}

\smallskip

The key point of our approach is that the Hamiltonian, $H_\text{eff}=\sqrt{12\pi G}K_y$, is simply an element of the $\su(1,1)$ algebra. Then the evolution is just given by the $\SU(1,1)$ transformations generated by $K_y$. This simple structure will furthermore hold at the quantum level when properly quantizing the $\su(1,1)$ structure without anomaly.

A remark is that while the observable $e^{\pm i b}$ corresponds to the holonomy variable in the context of loop quantum gravity/cosmology, our new observables $ve^{\pm i b}$  have a similar interpretation as a $T^1$-loop, that is a holonomy around a closed loop with one triad insertion (see e.g. \cite{ashtekar_lqg} for a review of the loop algebra underlying loop quantum gravity).

\subsection{Integrating the Equations of Motion as $\SU(1,1)$ Transformations}
\label{motion}

To compute the evolution, we can use the fundamental representation of $\SU(1,1)$ in terms of 2$\times$2 matrices with the algebra generators given by the Lorentzian Pauli matrices (see appendix \ref{AppA} for more details). Then the evolution is given by $\SU(1,1)$ transformations $e^{i\phi H_\text{eff}}=e^{i\tau\sigma_y}\equiv U_\tau$, where we have used $\tau=\sqrt{12\pi G}\phi$.
Computing the exponential we get
\begin{align}
U_\tau=\left( \begin{array}{cc}
\cosh\frac{\tau}{2} & \sinh\frac{\tau}{2} \\
\sinh\frac{\tau}{2} & \cosh\frac{\tau}{2}\end{array} \right).
\end{align}
Then we can derive the trajectories for $J_z$, $K_\pm$, or equivalently for $v$ and $b$, by acting on the matrix $M$ and computing $M(\tau)\equiv U_\tau M(0)U_\tau^\dagger$.
While the matrix $M$ lives in the adjoint representation, it is possible to introduce spinorial variables that live in the fundamental representation. It is much easier to integrate the equation of motion in these variables and they will also be more convenient when defining and studying coherent states at the quantum level.

\smallskip

As explained in more details in appendix \ref{SU11spinor}, from canonical complex variables $z^{0,1}$, one gets a representation of the $\su(1,1)$ algebra:
\be
\label{defJJ}
J_z=\f12\,\left(|z^0|^2+|z^1|^2\right),
\qquad
K_+=\bz^0\bz^1,
\qquad
K_-=K_+^\dagger=z^0z^1\,.
\ee
The advantage of this approach is that the spinor $z$, with components $z^0$ and $\bar z^1$ now  lives in the fundamental  representation of $\SU(1,1)$, namely $\SU(1,1)$ transformations represented as $2\times 2$ matrices act on $z$ simply by matrix multiplication.

From the 3-vector $(K_x,K_y,J_z)\in\R^3$, one can reconstruct uniquely the spinor $z\in\C^2$ up to a global phase. The only constraint is that the Casimir $C=\vJ^2$ has to be positive or equal to 0. As shown in the previous section, our loop cosmology phase space has a vanishing $\su(1,1)$ Casimir and can thus be recast in these terms. Solving for $J_z=v$ and $K_+=v e^{+ib}$, we easily get:
\be
z^0=\sqrt{v}\,e^{-i\f b2}\,e^{i\vphi},\quad
z^1=\sqrt{v}\,e^{-i\f b2}\,e^{-i\vphi}\,,
\ee
where $\vphi$ is an arbitrary phase. Due to the square-roots, we see that the constraint that the volume $v$ is positive, $v\ge 0$, is directly encoded at the kinematical level in the phase space structure defined in terms of these spinorial variables. These definitions can be generalized to the case of a non-vanishing $\su(1,1)$ Casimir $C>0$ and we will see later that it corresponds to the existence of a non-zero minimal volume.

Then, starting with an initial spinor $z(0)$ at $\tau=0$, the evolution is simply given by:
$$z(\tau)=\left( \begin{array}{c}
z^0(\tau) \\ \bar z^1(\tau)\end{array} \right)=U_\tau \triangleright z(0)=U_\tau  z(0).$$
From the expressions of $J_z$ and $K_x$ in terms of $z$, we deduce their evolution:
\begin{align}\label{evol}
J_z(\tau)=J_z(0)\cosh\tau+ K_x(0)\sinh\tau\,,\qquad K_x(\tau)=J_z(0)\sinh\tau+K_x(0)\cosh\tau\,.
\end{align}
As expected, this is simply the action on the 3-vector $(K_x,K_y,J_z)$ of the pure boost $U_\tau=e^{i\tau \sigma_y}$ in the $(z,x)$ plane.
We can re-absorb the initial conditions $J_z(0)$ and $K_x(0)$ in a different origin point for the time:
\be\label{evolution}
J_z(\tau)=J_z[z(\tau_o)]\cosh (\tau-\tau_o)\,,\qquad K_x(\tau)=J_z[z(\tau_o)]\sinh (\tau-\tau_o).
\ee
The component $K_x$ vanish at $\tau_o$ while $J_z$ reaches its minimal value.
Converting back into our standard cosmological variables, using the definitions $J_z=v$ and $K_x=v\cos b$, we obtain
\begin{align}
v(\tau)=v_o\cosh(\tau-\tau_o)\,,\qquad \cos b(\tau)=\tanh(\tau-\tau_o)\,.
\end{align}
As it should be, these trajectories coincide with the ones previously given in \eqref{sol} and the time origin $\tau_o$ corresponds to the minimal value of the volume and to the cosmological bounce.

\smallskip

Through this analysis, we see that the simple hyperbolic trajectories for the volume $v$ is somehow due to the ``hidden" $\su(1,1)$ structure of our space of observables and to the fact that the Hamiltonian is simply a boost generator in this framework.

\section{Group Structure of the Fully Regularized FRW Cosmology}
\label{Sec3}

The above description of the effective dynamics underlying LQC only takes into account the regularization of the variable $b$.
However, LQC also introduces a regularization of the volume as a consequence of a superselection of the kinematical Hilbert space. Let us be more explicit.
In LQC the geometry sector of the kinematical Hilbert space, ${\mathcal H}_\text{g}$, turns out to be Bohr compactification of the real line \cite{abl,Vel}. In momentum representation, and denoting by $|\nu\rangle$ the basis states,\footnote{According with our convention when defining $v$, we have $\nu=2v$, being $\nu$ the usual volume variable employed in the LQC literature.} then ${\mathcal H}_\text{g}$ is the completion of the space spanned by the states $|\nu\in\mathbb{R}\rangle$ in the discrete norm $\langle \nu'|\nu\rangle=\delta_{\nu',\nu}$ (here $\delta_{\nu',\nu}$ denotes the Kronecker delta). It turns out that ${\mathcal H}_\text{g}$ can be written as the direct sum of an infinite number of superselected sectors: ${\mathcal H}_\text{g}=\oplus_{\varepsilon} \mathcal H_{\varepsilon}^+\oplus \mathcal  H_{\varepsilon}^-$, where each sector $\mathcal H_{\varepsilon}^\pm$ is the space spanned by the states $|\nu\rangle$ with support in the lattices of constant step $\mathcal L_{\varepsilon}^\pm=\{\pm(\varepsilon+4n), n\in\mathbb{N},\varepsilon\in(0,4]\}$ (see e.g. \cite{mop}).
The different values of $\pm\varepsilon$ label inequivalent quantum theories in which the physical Hilbert space turns out to be precisely any of the superselection sectors.
In conclusion, the momentum space is spanned by the states  $|\nu=\pm\varepsilon\pm 4n\rangle$, so that $|\nu|$ displays a minimum value $\varepsilon$.

\medskip

Our description of the regularized phase space using the $\SU(1,1)$ structure can be easily generalized to account for the existence of a non-zero minimal volume at kinematical level, $v\ge v_m>0$. This is achieved through a simple regularization of the volume, roughly switching $v$ by $\sqrt{v^2-v_m^2}$. This can be taken into account by a basic modification of the set of observables. We know define:
\begin{align}\label{obs}
J_z=v, \qquad K_+=\sqrt{v^2-v_m^2} e^{ib},\qquad K_-=\sqrt{v^2-v_m^2} e^{-ib},
\end{align}
with fixed $v_m>0$.
As easily checked, these observables still form a $\su(1,1)$ algebra, and obviously assume that $v\ge v_m$. Considering these observables as fundamental, we can invert this definition and compute $v$ and $b$ in terms of the 3-vector $(K_x,K_y,J_z)$:
\be
v=J_z,
\qquad
\sin b= \f{K_y}{\sqrt{v^2-v_m^2}},
\qquad
\cos b= \f{K_x}{\sqrt{v^2-v_m^2}}\,.
\ee
The norm of the 3-vector defines the $\su(1,1)$ Casimir, which is now strictly positive, $C=\vJ^2=v_m^2>0$.

\smallskip

Similarly as before, we choose the $K_y$ boost generator as our effective Hamiltonian driving the dynamics of this fully regularized model:
\be
p_\phi=\tilde H_\text{eff}\,\equiv\,\sqrt{12\pi G}\sqrt{v^2-v_m^2}\,\sin b=\sqrt{12\pi G}\,K_y\,,
\ee
which takes directly into account the minimal kinematical volume $v_m$.
Evolution will be given by $\SU(1,1)$ transformations on our initial data and will always respect the kinematical constraint $v\ge v_m$, which has been encoded into the observables and the dynamics of the model.

\smallskip

To integrate the equations of motion as $\SU(1,1)$ transformations, it is  convenient to introduce the spinor variable as before. From their definition \eqref{defJJ} and the new expressions for the $\su(1,1)$ generators, we define:
\be
z^0=\sqrt{v+v_m}\,e^{-i\f b2}\,e^{i\vphi},\quad
z^1=\sqrt{v-v_m}\,e^{-i\f b2}\,e^{-i\vphi}\,,
\ee
where $\vphi$ is once again an irrelevant arbitrary phase and where the modulus of the two spinor components are now slightly different and depend on the value of the minimal kinematical volume.
Actually, one defines the following quantity:
\be
L(z)\equiv\,\f12(|z_0|^2-|z_1|^2)
\,=\,v_m>0.
\ee
The spinor $z=(z^0,\bz^1)$ then lives in the fundamental representation of $\SU(1,1)$. It transforms as $z\arr U\rhd z= U\,z$ for $\SU(1,1)$ and the spinor pseudo-norm  $L(z)$  actually turns out to be a $\SU(1,1)$-invariant.

Then we get the evolution of our spinor by simply acting on it with our evolution matrix $U_\tau\in\SU(1,1)$.
From there, computing the evolution of the various variables is rather direct and we obtain
\begin{align}\label{sol2}
v(\tau)=v_o\cosh(\tau-\tau_o)\,,\qquad \cos b(\tau)=\frac{v_o\sinh(\tau-\tau_o)}{\sqrt{v^2(\tau)-v_m^2}}\,,
\end{align}
where the volume $v_o$ at the bounce point $\tau_o$ is determined by the parameters $p_\phi$ and $v_m$ of this fully regularized model via
\be\label{vol-bounce}
v_o=\sqrt{\frac{p_\phi^2}{12 \pi G}+v_m^2}\quad\ge v_m\,.
\ee
This dynamical minimal volume depends on the actual trajectory through the parameter $p_\phi$, but is always larger than the postulated kinematical minimal volume.
Moreover, we see that only the trajectory for $b$ is modified, but as before the low curvature regime $b\arr 0$ is reached as $\tau$ grows to $\pm\infty$, so that this regularized dynamics is in agreement with general relativity far away from the bounce.

The matter density now reads
\begin{align}\label{dens}
\rho_\phi(\tau)=\frac{3}{8\pi G \lambda^2}\cdot\frac{p_\phi^2}{p_\phi^2+12 \pi Gv_m^2}\cdot\frac{1}{\cosh^2(\tau-\tau_o)}\,.
\end{align}
As before, it reaches a non-divergent maximum at the bounce:
\begin{align}
\rho_\phi(\tau_o)=\frac{3}{8\pi G \lambda^2}\cdot\frac{p_\phi^2}{p_\phi^2+12 \pi Gv_m^2}
\,\underset{p_\phi\gg v_m}\sim\, \frac{3}{8\pi G \lambda^2}\,.
\end{align}
This value now depends in general on the value of the momentum of the scalar field, though this dependence is negligible in the case $p_\phi^2\gg 12 \pi Gv_m^2$. Moreover, in this regime the maximum is of Planck order.
Note that in the case $v_m=0$ we recover the effective dynamics of previous sections.


\section{Loop Quantum FRW Cosmology by Group Theoretical Quantization}
\label{Sec4}

\subsection{Quantizing the Effective Dynamics and Super-Selection Sectors}

\label{quant}

Now that we have made explicit the $\su(1,1)$ structure of the (fully) regularized phase space of the FRW model coupled to a massless scalar within LQC, we can quantize the model simply by considering the irreducible representations of the group $\SU(1,1)$.

As we have seen before, in our model the Casimir is positive, and therefore among all the irreducible representations of $\SU(1,1)$ we are interested in those of the discrete principal series (time-like representations). In order to derive them we employ the spinor formulation of the $\su(1,1)$ algebra, as described in appendix \ref{A2}. Writing the $\su(1,1)$ generators in terms of the canonical complex variables $z^{0,1}$ as introduced above in \eqref{defJJ}, we quantize the system as a pair of harmonic oscillators: raising $z^{0,1}$ to annihilation operators while their complex conjugate $\bz^{0,1}$ become creation operators. The $\su(1,1)$ generators $J_z$ and $K_\pm$ are then quadratic in those basic operators. We obtain a hugely reducible representation of $\SU(1,1)$. It is however easily realized that the $\SU(1,1)$ Casimir depends on the difference of energy between the two oscillators. Fixing this energy, we finally obtain the whole discrete principal series of $\SU(1,1)$ representations.
The derivation of these representations has been detailed in appendix \ref{B1}. Let us summarize here their main properties.

The generators $J_z$, $K_x$ and $K_y$ are promoted to Hermitian operators satisfying the $\su(1,1)$ commutation relations, or equivalently expressed in terms of $J_z$ and $K_\pm$:
\be
[J_z,K_\pm]=\pm K_\pm,\qquad
[K_+,K_-]=-2J_z,\qquad
J_z^\dagger=J_z,\quad
K_-=K_+^\dagger\,.
\ee
To characterize the irreducible representations, one diagonalizes the operator $L$, related with the Casimir operator
$
C=J_z^2-\frac12(K_+K_-+K_-K_+)
$
through the expression
\be
C=\left(L+\f12\right)\left(L-\f12\right).
\ee
In our method, $L$ has a discrete spectrum (see appendix \ref{B1} for more details) and we are exploring the eigenvalues of the Casimir operators belonging to its discrete spectrum. Consequently the eigenvalues of $L$ label the time-like irreducible representations of $\SU(1,1)$.

We use the usual $\SU(1,1)$  basis diagonalizing both the Casimir and the operator $J_z$. The basis states are then labeled by a spin $j$ giving the eigenvalue of $L$ and by the magnetic moment giving the value of $J_z$. The action of the $\su(1,1)$ generators on this orthonormal basis is:
\beq
L|j,m\ra&=&\left(j-\f12\right)|j,m\ra\,,
\\
C|j,m\ra&=&j(j-1)|j,m\ra\,,\nn\\
J_z|j,m\ra
&=&
m|j,m\ra\,,\\\nn
K_+|j,m\ra
&=&
\sqrt{(m-j+1)(m+j)}|j,m+1\ra\,,\nn\\
K_-|j,m\ra
&=&
\sqrt{(m-j)(m+j-1)}|j,m-1\ra\,.\nn
\eeq
We obtain two types of representations: the discrete positive series with $\f12\leq j\leq m=j+\mathbb{N}$; and the negative one with $-\f12\ge j\ge m=j-\mathbb{N}$. In both cases $|j|$ is any positive half integer.
Thus the irreducible representations of spin $j\ge\f12$ live on the Hilbert spaces spanned by the basis states $|j,m\ra$ with $m\ge j$, $\cV_+^j\equiv\bigoplus_{m\ge j} \C\,|j,m\ra$, and the irreducible representations of spin $j\leq-\f12$ live on the Hilbert spaces dual to the previous ones, $\cV_-^j\equiv\bigoplus_{m\le -j} \C\,|j,m\ra=\overline{\cV_+^j}$.
As it is usually done in LQC, we will restrict our study to the sector with positive eigenvalues of $J_z=v$,  namely we will only consider the irreducible representations of positive spin.

We see that as in usual LQC, the kinematical volume, in our case denoted by $m$, presents a minimum $m=j$ that labels the different superselection sectors. In our case this minimum turns out to be discretized in the quantum theory since it takes values in the discrete set
$\f\N2+\f12$.
Then our approach features superselection sectors labeled by a countable parameter.

\subsection{Quantum Evolution and Coherent Wave-Packets}
\label{coherent}

Let us now look into the dynamics at the quantum level. Our Hamiltonian is the boost generator $K_y$. As is well-known its spectrum is the whole real line and we can construct its eigenvectors in the considered representation (see appendix \ref{Ky} for more details). We would like however to focus here on the construction of coherent states, with good semi-classical properties and whose shape is preserved under evolution.

One of the beauties of our $\SU(1,1)$ quantization consists in the simplicity of providing such states that are coherent under evolution.
In fact, the evolution operators, given by $U_\tau=e^{i\tau K_y}$, are  $\SU(1,1)$ group elements. Therefore $\SU(1,1)$ coherent states provide dynamical coherent states.

As we define and review in appendix \ref{AppB}, $\SU(1,1)$ coherent states in the $\cV_+^j$ representation, have the explicit expression
\be
|j,z\ra\,\equiv\,
\sum_{m\in j+\N}^\infty \sqrt{\f{(m+j+1)!}{(m-j)!(2j+1)!}}\,
\left(\f1{\bz^0}\right)^{m+j+2}\,(z^1)^{m-j}\,|j,m\ra\,,
\ee
%
These coherent states are labeled by a classical spinor $z\in\C^2$, whose components $z^0$ and $\bar z^1$ are arbitrary complex numbers verifying $|z^0|>|z^1|$.
They provide an over-complete basis for the physical Hilbert space $\cV_+^j$, as we show in section \ref{B3}.

Their key property is that they transform covariantly under $\SU(1,1)$, i.e the action of $\SU(1,1)$ transformations on those states act directly on their label:
\be
U\,|j,z\ra\,=\,|j,\,U\vartriangleright z\ra \,=\,|j,\,U\,z\ra\,.
\ee
Thanks to this, it is straightforward to compute their quantum evolution. The coherent states will follow the classical trajectory computed earlier as a $\SU(1,1)$ flow. Only their fluctuations around the classical expectation value will evolve.
Explicitly, starting from an initial state $|j,z(0)\ra$ characterized by the initial conditions $z^{0,1}(0)$ given at some initial time $\tau=0$, the evolved state at time $\tau$ is simply given by $|j,z(\tau)\ra=|j, U_\tau\vartriangleright z(0)\ra$.
Here
\begin{align}
U_\tau=\left( \begin{array}{cc}
\cosh\frac{\tau}{2} & \sinh\frac{\tau}{2} \\
\sinh\frac{\tau}{2} & \cosh\frac{\tau}{2}\end{array} \right),\quad \text{and}\quad
z(0)=\left( \begin{array}{c} z^0(0) \\ \bar{z}^1(0) \end{array}\right),
\end{align}
therefore the evolved coherent state  $|j,z(\tau)\ra$ is labeled by the spinor
\begin{align}\label{ztau}
z(\tau)=\left( \begin{array}{c} z^0(\tau) \\ \bar{z}^1(\tau) \end{array}\right)
=\left( \begin{array}{c}
z^0(0)\cosh\frac{\tau}{2} + \bar{z}^1(0) \sinh\frac{\tau}{2}\\
z^0(0) \sinh\frac{\tau}{2}+\bar{z}^1(0)\cosh\frac{\tau}{2}
\end{array}\right).
\end{align}
Now we only have to specify a suitable initial spinor $z(\tau=0)$ and we will have the whole quantum evolution described in terms of coherent states.

\smallskip

The physical meaning of these coherent states is given by the expectation values of physical observables, which determine on which phase space point these states are peaked, and by the fluctuations of the observables, which determine how semi-classical the states are.
As computed in appendix \ref{B2}, the expectation values of the $\su(1,1)$ generators are:
\be
\la \vJ\ra
\,\equiv\,
\f{\la j,z|\vJ|j,z\ra}{\la j,z|j,z\ra}
\,=\,
j\f{\vJ(z)}{L(z)}\,,
\ee
where we remind the definition of the $\SU(1,1)$ invariant $L(z)\equiv\,(|z_0|^2-|z_1|^2)/2>0$ and the coherent state norm is shown to be $\la j,z|j,z\ra=1/(2L(z))^{2j}$.
Following the analysis of the classical case presented in section \ref{Sec3}, the norm of these expectation values $\vJ$ as a 3-vector gives the value of the minimal kinematical volume:
\be
v_m^2=\la \vJ\ra^2= j^2\f{\vJ(z)^2}{L(z)^2}=j^2
\quad \Longrightarrow
v_m=j\,,
\ee
as we expect since the minimal kinematical volume at the quantum level in the irreducible representation of spin $j$ is actually the spin $j$ itself by definition of the Hilbert space.
Then let us notice that the expectation values $\la \vJ\ra$ are invariant under rescaling of the spinor $z$. This is actually a property of the coherent states themselves (see in appendix for more details). We are thus free to fix the pseudo-norm $L(z)$ of the spinor as we want. For convenience, in order to match the description of the classical dynamics in terms of spinors and to get rid of the normalization factors $j/L(z)$, we will fix without loss of generality:
\be
L(z)=v_m=j\,.
\ee

Now, in order to extract the meaning of these expectation  values, we focus on the complete set of physical observables formed by the volume operator $V=4\pi G\lambda J_z$ and by the momentum of the scalar field $p_\phi=\sqrt{12\pi G} K_y$. Since both observables are $\su(1,1)$ elements, we already have their expectation values from the formula above. We can start with a coherent state labeled by the spinor $z(0)$ peaked on a fixed value of $p_\phi$ and a given arbitrary volume $V$. Then we evolve this initial quantum state with $U_\tau=e^{i\tau K_y}\in\SU(1,1)$. This leads to the coherent state $|j,z(\tau)\ra$ with the spinor $z(\tau)$ given explicitly by:
\be
z^0(\tau)=\sqrt{v(\tau)+v_m}\,e^{-ib(\tau)/2}\,e^{i\varphi}\,,\qquad z^1(\tau)=\sqrt{v(\tau)-v_m}\,e^{-ib(\tau)/2}\,e^{-i\varphi},
\ee
with $v(\tau)$ and $b(\tau)$ determined in \eqref{sol2} and \eqref{vol-bounce}. Here $\varphi$ is an arbitrary irrelevant phase, which neither evolves nor affects the physical expectation values of our observables. Checking the expectation values on these coherent states $|j,z(\tau)\ra$, we get as wanted:
\be
\la \vJ\ra
\,=\,
\left(v(\tau)\cos b(\tau),v(\tau)\sin b(\tau),v(\tau)\right),\quad
\Rightarrow
\la p_\phi\ra=p_\phi,
\quad
\la V(\tau) \ra=  V(\tau)=
4\pi G\lambda \sqrt{\frac{p_\phi^2}{12 \pi G}+v_m^2}\,\cosh(\tau-\tau_o)\,.
\ee
The expectation values of the physical observables follow exactly the classical trajectory. First note that the field momentum $p_\phi$ is of course a constant of motion. Then the behavior of  the volume confirms that the universe undergoes a quantum bounce at $\tau=\tau_o$, and that this bounce is universal  regardless of the particular values of the spinor components labeling the coherent states.

\smallskip

Next we would like to check the quantum fluctuations around the classical trajectory. Since our physical  observables -volume and momentum- are $\su(1,1)$ generators, we also know their uncertainties (see appendix \ref{B2}):
\beq
&&\f{\Delta p_\phi}{\la p_\phi \ra}
\,=\,\f{\sqrt{\la K_y^2\ra-\la K_y\ra^2}}{\la K_y\ra}
\,=\,\sqrt{\f{1}{2v_m}\left(1+\f{12\pi G v_m^2}{p_\phi^2}\right)} \\
&&\f{\Delta V}{\la V\ra}(\tau)
\,=\,\frac{\sqrt{\la J_z^ 2\ra-\la J_z\ra^2}}{\la J_z \ra}
\,=\,\sqrt{\frac{1}{2v_m}\left[1-\left(\f{4\pi G\lambda v_m}{V(\tau)}\right)^2\right]}\nn
\eeq
We see that the evolution of the expectation value and fluctuation of the volume are symmetric around the bounce. Moreover, keeping in mind that $v_m=j$, the relative fluctuation varies from a minimum value at the bounce to a maximum value equal to $\sqrt{1/(2j)}$, approached in the limits $\tau\rightarrow\pm\infty$ where the expectation value of the volume tends to infinity. This explicitly shows that the relative fluctuation in the volume displays a universal (state-independent) bound that only depends on the representation. Furthermore, the larger $j$ is, the smaller the fluctuations are, both in the volume and in the momentum of the field.

In conclusion, the $\SU(1,1)$ coherent states, that we propose here, provide good coherent semiclassical states for the Hamiltonian $H[v,b]=\sqrt{v^2-v_m^2}\sin b=p_\phi/\sqrt{12\pi G}$ for arbitrary values of the parameters $p_\phi$ and $v_m$.

We can also  look at the evolution of the matter density $\rho_\phi=p_\phi^2/(2V^2)=3K_y^2/(8\pi G\lambda^2 J_z^2)$. However the matter density operator is not as neat as the volume operator. It is not linear in the $\su(1,1)$ generators and it will thus suffer from factor ordering ambiguities since $J_z$ and $K_y$ do not commute. Moreover its exact action on (coherent) states is a priori not obvious. Nevertheless, since our coherent states are semiclassical and properly peaked on the classical trajectory, we can approximate the expectation value of the density on these states by its classical value:
\be
\la \rho_\phi(\tau)\ra\approx \f{3}{8\pi G\lambda^2}\f{\la K_y^2\ra}{\la J_z^2 \ra}=\rho_\phi(\tau)\,,
\ee
given in \eqref{dens}.
Provided that the relative fluctuations of $\rho_\phi$, $p_\phi$ and $V$ are small, we can do the following approximation to compute the relative fluctuation of the matter density on coherent states:
\be
\f{\Delta\rho_\phi}{\la \rho_\phi \ra}\approx \f{\delta\rho_\phi}{\rho_\phi}=2\left[\f{\delta p_\phi}{p_\phi}-\f{\delta V}{V}\right]
\approx 2\left[\f{\Delta p_\phi}{\la p_\phi \ra}-\f{\Delta V}{\la V\ra}\right]\,.
\ee
The result is
\be
\f{\Delta\rho_\phi}{\la \rho_\phi \ra}(\tau)\approx \sqrt{\f2{v_m}}
\left[\sqrt{\f{p_\phi^2+12\pi G v_m^2}{p_\phi^2}}-\sqrt{1-\f{12\pi G v_m^2}{p_\phi^2+12\pi G v_m^2}\cdot\f{1}{\cosh^2(\tau-\tau_o)}}\right]\,.
\ee
This fluctuation reaches its maximum at the bounce:
\be
\f{\Delta\rho_\phi}{\la \rho_\phi \ra}(\tau_o)\approx \sqrt{\f2{v_m}}
\left[\sqrt{\f{p_\phi^2+12\pi G v_m^2}{p_\phi^2}}-\sqrt{\f{p_\phi^2}{p_\phi^2+12\pi G v_m^2}}\right]\,,
\ee
and tends to a minimum value in the limits $\tau\rightarrow\pm\infty$:
\be
\f{\Delta\rho_\phi}{\la \rho_\phi \ra}\xrightarrow{\tau\rightarrow\pm\infty} \sqrt{\f2{v_m}}
\left[\sqrt{\f{p_\phi^2+12\pi G v_m^2}{p_\phi^2}}-1\right]\,.
\ee
For large values of the field momentum  $p_\phi^2\gg 12\pi G v_m^2$, the coherent states will have minimal spread in $p_\phi$ and the uncertainty on the matter density will vanish.

Finally we would like to point out that our analysis is  valid for strictly positive values of the minimal volume $v_m>0$ and we do not have well-defined coherent states in the special case of vanishing $v_m$.

\subsection{Comparison with Other LQC Hamiltonians and Operator Orderings}

Here, we have identified and discussed the $\SU(1,1)$ structure of the flat FRW model in loop cosmology. The group theoretical quantization ensures that the $\SU(1,1)$ structure is preserved at the quantum level without anomaly. This fixes all the ordering ambiguities appearing in the definition of the $\su(1,1)$ operators at the quantum level and in particular entirely determines the Hamiltonian operator. This is an improvement with respect to the usual LQC context, where we find various proposals for the Hamiltonian constraint operator, whose exact behaviors at small scales are different.

To summarize our proposal, our Hilbert space is any of the time-like irreducible representations of the group $\SU(1,1)$ with spin $j$ positive (for instance): $\cV_+^j$.
A basis for this Hilbert space is provided by the eigenstates $|j,m\rangle$ (with $m=j+\mathbb{N}$) of the Casimir $L$, defined such that $L|j,m\rangle=(j-1/2)|j,m\rangle$. Our basic operators are also $J_z$ representing the volume $v$, with diagonal action on the basis states: $J_z |j,m\rangle=m|j,m\rangle$, and $K_\pm$ representing the observables $\sqrt{v^2-v_m^2}e^{\pm i b}$, with action on basis states given by
\be
K_+|j,m\ra=\sqrt{(m-j+1)(m+j)}|j,m+1\ra\,,\qquad
K_-|j,m\ra=\sqrt{(m-j)(m+j-1)}|j,m-1\ra\,.
\ee
Moreover, the Hamiltonian operator is
\be
H=\sqrt{12\pi G}K_y=\f{\sqrt{12\pi G}}{2i}(K_+-K_-).
\ee
In LQC the set of basic operators are $v$ and the holonomies $e^{\pm i b}$, which act on the
volume eigenstates
by translation of $\pm 1$ units respectively. We would like to note that in our approach although those holonomies are not basic operators, they can be defined as well, in the following way:
\be
\widehat{e^{ib}}:=\f1{\sqrt{J_z-L-\f12}}K_+ \f1{\sqrt{J_z+L+\f12}}\,,\qquad
\widehat{e^{-ib}}=\widehat{e^{ib}}^\dagger=\f1{\sqrt{J_z+L+\f12}}K_- \f1{\sqrt{J_z-L-\f12}}\,.
\ee
We recover the expected action $\widehat{e^{\pm ib}}|j,m\ra=|j,m\pm 1\ra$, thanks to the non-trivial regularization of the inverse volume $1/v=1/J_z$.

We can square our Hamiltonian operator to get the gravitational part of the Hamiltonian constraint operator, that we will denote by $\Theta(j)$. Let us note that its action on the basis states $|j,m\rangle$ is given by
\begin{align}
\Theta(j)|j,m\ra&=-3\pi G\left[\sqrt{(m+1-j)(m+j)}\sqrt{(m+2-j)(m+1+j)}|j,m+2\ra-2(m^2-j^2+j)|j,m\right.\ra\nn\\
&\left.+\sqrt{(m-1+j)(m-j)}\sqrt{(m-2+j)(m-1-j)}|j,m-2\ra\right]\,.
\end{align}
Let us compare it with the various proposals for FRW in LQC. In concrete we will consider the {\it Ashtekar-Paw{\l}owski-Singh} prescription (APS) \cite{aps3}, the {\it solvable LQC} prescription (sLQC) \cite{acs}, the {\it Mart\'{i}n--Benito-Mena  Marug\'{a}n-Olmedo} prescription (MMO) \cite{mmo}, and the {\it solvable MMO} prescription (sMMO) \cite{mop}.
For a comparison between them we refer the reader to \cite{mop}. We summarize here how the Hilbert space and the geometry term of the Hamiltonian constraint operator look like in each case. In all these cases the Hilbert space is either $\mathcal H^\pm_\varepsilon$, the space spanned by the basis states $|\nu\ra$ with $\nu\in\mathcal L^\pm_\varepsilon:=\{\pm(\varepsilon+4n),n\in\mathbb{N}\}$ and normalizable with respect to the discrete inner product, or
$\mathcal H_\varepsilon=\mathcal H^+_\varepsilon\oplus \mathcal H^-_{4-\varepsilon}$. Generically, the geometry term of the Hamiltonian constraint operator reads
\be
\Theta|\nu\ra=-\f{3\pi G}{4}\left[f(\nu+2)|\nu+4\ra-f_o(\nu)|\nu\ra+f(\nu-2)|\nu-4\ra \right]\,.
\ee
Each prescription is characterized by the specific form of the functions $f(\nu)$ and $f_o(\nu)$, though all of them agree in the large $\nu$ limit. Moreover, in all the cases the operator is essentially self-adjoint \cite{kale}:
\begin{itemize}
\item APS:
\begin{align}
f(\nu)&=\sqrt{\beta(\nu+2)\beta(\nu-2)}|\nu|\big||\nu+1|-|\nu-1|\big|\,,\\
f_o(\nu)&=\beta(\nu)\left[(1-\delta_{\nu,-4})|\nu+2|\big||\nu+3|-|\nu+1|\big|+
(1-\delta_{\nu,4})|\nu-2|\big||\nu-1|-|\nu-3|\big|\right]\,,\nn\\
\beta(\nu)&=\begin{cases} \f{4}{27|\nu|} \big||\nu+1|^\f13-|\nu-1|^\f13\big|^{-3}& {\text{if}} \quad \nu\neq 0,\\
0 & {\text{if}} \quad \nu=0.\nn\\
\end{cases}\\
\end{align}
This prescription is defined in the Hilbert space $\mathcal H_\varepsilon$, namely it does not decouple the semi-axis $\nu>0$ from the semi-axis $\nu<0$.
\item sLQC:
\be
f(\nu)=|\nu|\sqrt{|\nu+2||\nu-2|}\,,\qquad f_o(\nu)=2\nu^2\,.
\ee
This prescription does not involve corrections coming from the inverse volume operator and then it is simpler and indeed leads to an analytically solvable quantum model. As the previous prescription, negative and positive semi-axis are not decoupled, and then the Hilbert space is $\mathcal H_\varepsilon$.
\item MMO:
\begin{align}
f(\nu)&=\f1{9} g(\nu+2)g(\nu-2)g^2(\nu)s_+(\nu)s_-(\nu)\,,\\
f_o(\nu)&=\f1{9}g^2(\nu)\left\{[ g(\nu+2)s_+(\nu)]^2+[ g(\nu-2)s_-(\nu)]^2\right\}\,,\\
g(\nu)&=\begin{cases}
\left|\left|1+\frac1{\nu}\right|^{\frac1{3}}
-\left|1-\frac1{\nu}\right| ^{\frac1{3}}
\right|^{-\frac1{2}} & {\text{if}} \quad \nu\neq 0,\\
0 & {\text{if}} \quad \nu=0\\
\end{cases}\,,\qquad s_\pm(\nu)=\text{sign}(\nu\pm2)+\text{sign}(\nu).\nn
\end{align}
This prescription takes into account the sign of $\nu$ in the factor ordering to explicitly decouple the negative semiaxis from the positive one. The Hilbert space is then either $\mathcal H^-_\varepsilon$ or $\mathcal H^+_\varepsilon$. Usually one just considers $\mathcal H^+_\varepsilon$.
\item sMMO:
\be
f(\nu)=\f14|\nu|\sqrt{|\nu+2|\nu-2|}s_+(\nu)s_-(\nu)\,,\qquad f_o(\nu)\f14|v|\big[|\nu+2|s_+^2(\nu)+|\nu-2|s_-^2(\nu)\big]\,.
\ee
This prescription combines the simplicity of sLQC with the decoupling of MMO of the positive semiaxis from the negative one, then one can define the operator $\Theta$ in the Hilbert space $\mathcal H^+_\varepsilon$ for which $\nu>0$.
\end{itemize}

It is straightforward to realize that our operator for the irreducible representation $j=1$ exactly matches the sMMO prescription (for the sector $\varepsilon=2$ strictly speaking).\footnote{We remind that according with our conventions $\nu=2m$.} This hints that the deep reason behind the solvability of sLQC  (apparently unnoticed by its authors) is in fact its $\SU(1,1)$ structure, which automatically provides a self-adjoint representation for the Hamiltonian and allows to fully and exactly integrate the evolution at the quantum level.

\smallskip

Moreover, in standard LQC the parameter $\varepsilon$ labeling different superselection sectors does not have any classical counterpart.
Indeed, the resulting classical effective dynamics does not distinguish this minimum kinematical volume. In this sense, our approach is more general since at the classical level it allows to distinguish different values of the  minimum volume $v_m=j$. Then our proposed classical regularized (effective) dynamics is controlled by both $v_m$ and $\lambda$. In that sense, it is also cleaner since it automatically takes care of the restriction to the positive volume sector or more generally to the $v\ge v_m$ sector, which is now encoded directly in the definition of the classical phase space and of  Hilbert space by the simple requirement of working with an irreducible time-like representation of $\SU(1,1)$.

The $\SU(1,1)$ structure allows us to go further than previous analyzes when studying the quantum dynamics of the model. In fact, we have exact and closed formulas for truly dynamical coherent states and not merely semi-classical states.
As a result our work further clarifies whether the bounce preserves or not semi-classicality. This issue, sometimes called ``cosmic forgetfulness" or ``cosmic recall", has generated important discussions in the literature. We find results supporting that semi-classical states at any time remain semi-classical during the whole evolution (cosmic recall)\cite{kp-posL,CoM,cs,cs2}, or confronting results pointing out that states semi-classical at late times can have been highly quantum before the bounce (cosmic forgetfulness) \cite{boj3,boj4a,boj4b,boj4c}.

Among these studies, only Bojowald attempted to exploit the $\SU(1,1)$ structure in order to analyze the evolution of quantum states \cite{boj3}. He derived the equations of motion of the expectation values and variances of quantum states. Then imposing the condition of saturating the uncertainty relations, he obtained two classes of semi-classical states: a first family with bounded fluctuations before and after the bounce (recall states), to which our coherent states belong, and a second family of states semi-classical at late times but with important quantum fluctuations before the bounce (forgetful states).
%
Our coherent state construction confirms explicitly the existence of states from the first family, with good semi-classical properties before and after the bounce, and although our intuition is that there does not exist any forgetful coherent state saturating the uncertainty relations in our Hilbert space, our present analysis does not allow us to check such a claim.
We would like nevertheless to point out a major difference between our approach and  the construction introduced by Bojowald. Indeed a ``reality condition" $J_z^2-K_-K_+=0$ was imposed in \cite{boj3}. First, it looks very similar to fixing the $\SU(1,1)$ Casimir $J_z^2-(K_-K_++K_+K_-)/2$  to a vanishing value. This is precisely the case which we avoid in our construction, where we focus on time-like representations of $\SU(1,1)$ which correspond to strictly positive values of the Casimir. Thus the null-like case such as considered in \cite{boj3} might be qualitatively different. Coherent states for null-like representations are actually much subtler to construct (issue of vanishing norm states) and our definitions presented here can not be applied in a direct way.
Second the constraint $J_z^2-K_-K_+=0$ is not $\SU(1,1)$-invariant and therefore not preserved under evolution. Although it might finally turn out that changing this into a Casimir constraint does not affect the existence of recall/forgetful states as derived in \cite{boj3}, this seems to be a crucial ingredient of the definition of the quantum theory.

On the other hand, the obvious advantage of our approach is that we do build explicitly the coherent states minimizing the uncertainty relations and whose shape is stable under evolution. Therefore we always have under control the spread and expectation values of observables in the quantum states, which are explicitly and exactly computable. Our results confirm the universality of the quantum bounce and that relative fluctuations of the volume are bounded, confirming previous results on semi-classical states  \cite{kp-posL,CoM,cs,cs2}.

\section{Beyond the pure $K_y$ Hamiltonian}

\label{Sec5}

From the group theoretical perspective it is natural to wonder about the physical meaning of a generic Hamiltonian living in the $\su(1,1)$ algebra. It may happen that other $\su(1,1)$ elements have a geometrical interpretation such as a curvature term or a cosmological constant. We are going to investigate this possibility in this section.
The advantage of this method is that as long as the Hamiltonian is a $\su(1,1)$ Lie algebra element we can describe the evolution by finite $\SU(1,1)$ transformations and use the $\SU(1,1)$ coherent states to describe the semi-classical regime of the theory.
However, as soon as we depart from a $\su(1,1)$ Hamiltonian, the new interaction terms will induce evolution outside $\SU(1,1)$ and our set of coherent states will not be stable anymore under the dynamics.

\subsection{$\su(1,1)$ Hamiltonian with $K_x$ term}

Let us consider the introduction of a $K_x$ term in our $\su(1,1)$ Hamiltonian for the regularized model:
\be
H_{reg}=K_y+\alpha K_x,\qquad \alpha=\text{constant}\,.
\ee
Using $K_x=\sqrt{v^2-v_m^2}\cos b$ and $K_y=\sqrt{v^2-v_m^2}\sin b$ we can write
\be
H_{reg}=\sqrt{(1+\alpha^2)(v^2-v_m^2)} \sin \tilde b, \quad\text{with}\quad  \tilde b= b+\arccos\f1{\sqrt{1+\alpha^2}}\,.
\ee
We see that the $K_x$ term accounts for a displacement of the origin of the angle $b$ and a rescaling of the time variable, so that its effect on the evolution is
physically irrelevant.

\subsection{$\su(1,1)$ Hamiltonian with $J_z$ term}

Let us now consider the introduction of a $J_z$ term in our $\su(1,1)$ Hamiltonian, so that the dynamics for our regularized model is driven by
\be
H_{reg}=K_y+\beta J_z,\qquad \beta=\text{constant}\,.
\ee
Assuming that the observables are still given by $J_z=v$ and $K_x=\sqrt{v^2-v_m^2}\cos b$, the above Hamiltonian reads
\be
H_{reg}=\sqrt{v^2-v_m^2}\sin b +\beta v.
\ee
Let us compare it with the Hamiltonian of the FRW model coupled to a massless scalar field and with curvature or cosmological constant, reviewed in appendix \ref{AppC}. In the regime where we usually compare the effective dynamics for loop quantum cosmology to the classical FRW setting, for small $b\arr 0$ and $v>>v_m$, our effective Hamiltonian at leading order is $H_{reg}\sim\beta v$ and matches with the leading order of the Hamiltonian of the flat FRW model with negative cosmological constant for  $\beta\propto\sqrt{-\Lambda}$:
\be
H_{FRW}=v\sqrt{b^2-\f\Lambda 3} \sim v\sqrt{-\f\Lambda 3}.
\ee
However, this matching already breaks down for small fluctuations away from $b=0$ as we can see from computing the next to leading order:
$$
H_{reg}\underset{b\arr 0}{\sim} \beta v + vb
\qquad\ne\qquad
H_{FRW}\underset{b\arr 0}{\sim} v\sqrt{-\f\Lambda 3} + vb^2\,\f12\sqrt{-\f3\Lambda }\,,
$$
where the next-to-leading order in $b$ do not match
However this mismatch seems to be a particularity of the regime $b\arr 0$. For instance, for positive cosmological constant, $b$ can never reach 0, so this does not seem to be the correct regime in which to compare the FRW Hamiltonian and our regularized proposal. Let us consider a regime where $b$ would be peaked around some constant value up to small fluctuations.  Then the perturbations in $b$ would be controlled by the Hamiltonian expanded around arbitrary value $b_0$. Taking $b=b_0+\delta b$, we have:
\be
H_{reg}\underset{\delta b\arr 0}{\sim} (\beta+\sin b_0)v+\delta b v\cos b_0-v\delta b^2\f{\sin b_0}{2}
\qquad\overset{?}\longleftrightarrow\qquad
H_{FRW}\underset{\delta b\arr 0}{\sim} v\sqrt{\cB}
+v\delta b \f{b_0}{\sqrt{\cB}}
+ v\delta b^2\,\f1{2\sqrt{\cB}}\left(1-\f{b_0^2}{2\cB}\right)\,,
\ee
with $\cB=b_0^2-\Lambda /3$. These two expressions match (up to adjusting the various constants and potentially rescaling the volume $v$) and this hints towards a real possibility that the $J_z$ term in the $\su(1,1)$ Hamiltonian allows to take into account a non-vanishing cosmological constant. In order to check whether this could be true, let us look in more details at the equations of motion and the trajectories.

\smallskip

First we compute the equations of motion for our regularized dynamics:
\be
\pp_\tau v = -\sqrt{v^2-v_m^2}\cos b,
\qquad
\pp_\tau b = \beta +\f{v}{\sqrt{v^2-v_m^2}}\sin b\,.
\ee
Differentiating a second time the volume variable, we derive the corresponding Friedmann equation:
\be
\pp_\tau^2 v = (1-\beta^2)v+\beta H_{reg},
\ee
where $H_{reg}$ is obviously a constant of motion. Let us keep in mind that this is the Friedmann equation in the internal time (determined by the scalar field) and not in proper time. Nevertheless, comparing it with the Friedmann equation for non-zero cosmological constant as given in appendix \ref{C2}, we see that these crucially differ: the constant term $\beta H_{reg}$ is replaced by a term going in $\Lambda v^3$ with non-trivial scaling in the volume $v$. Therefore we should expect an important mismatch for large volume, while the local behavior around a volume extremum (maximal volume or bounce) would be similar.

\smallskip

We illustrate this by comparing explicitly the trajectories. For the classical FRW model, we keep the known trajectories in appendix \ref{C2}. For our regularized dynamics, the evolution is given by the finite $\SU(1,1)$ transformations defined by the group elements $U_\tau=e^{i\tau(K_y+\beta J_z)}$. Proceeding as before, we act with these 2$\times 2$ matrices on some initial spinorial data $z(0)$, from which we deduce $z(\tau)$ and thus the trajectories for $J_z(\tau)$ and $K_x(\tau)$ and finally for the geometric variables $v$ and $b$ using \eqref{defJJ}.

We distinguish two general different cases: $|\beta|<1$ and $|\beta|>1$, which correspond to the Hamiltonian being respectively a space-like of time-like vector in the Lie algebra $\su(1,1)$ identified to the $2+1$ Minkowski space-time $\R^{2,1}$. We will put aside the critical case $|\beta|=1$ and we will further restrict ourselves to $\beta>0$ for simplicity's sake.

\begin{itemize}
\item $0<\beta<1$: space-like Hamiltonian

In this case the evolution is given by the following boost transformation
\be
U_\tau=\left( \begin{array}{cc}
\cosh\f{\ttau}{2}+\f{i\beta}{\sqrt{1-\beta^2}}\sinh\f{\ttau}{2} & \f{1}{\sqrt{1-\beta^2}}\sinh\f{\ttau}{2}\vspace*{0.2cm} \\
\f{1}{\sqrt{1-\beta^2}}\sinh\f{\ttau}{2} &
\cosh\f{\ttau}{2}-\f{i\beta}{\sqrt{1-\beta^2}}\sinh\f{\ttau}{2}\end{array} \right),
\ee
where $\ttau\equiv\tau\,\sqrt{1-\beta^2}$.
Applying this matrix on suitable initial spinorial data and after a few straightforward algebraic manipulation, we get the resulting trajectories (see fig. \ref{hyper_plot}):
\be
v(\tau)
\,=\,
A\cosh(\ttau-\ttau_0)+B\sinh(\ttau-\ttau_0)-\f{\beta}{1-\beta^2}H_{reg}\,,
\ee
\be
\cos b(\tau)
\,=\,
\f{-\sqrt{1-\beta^2}}{\sqrt{v(\tau)^2-v_m^2}}\,
\left[A\sinh(\ttau-\ttau_0)+B\cosh(\ttau-\ttau_0)
\right]\,,
\ee
where the constants $A$ and $B$ depend on the initial conditions at $\ttau=\ttau_0$ and can be computed from the initial volume and the value of energy. It is easy to check that these satisfy the Friedmann equation and equations of motion given above.
%
%
Note that these trajectories reduce to those of section \ref{Sec3} for $\beta=0$, as expected.

The universe starts with infinite volume at $\tau\arr-\infty$, collapses, bounces and grows back to infinite volume at $\tau\arr+\infty$.

\begin{figure}[h]
\begin{center}
\includegraphics[height=35mm]{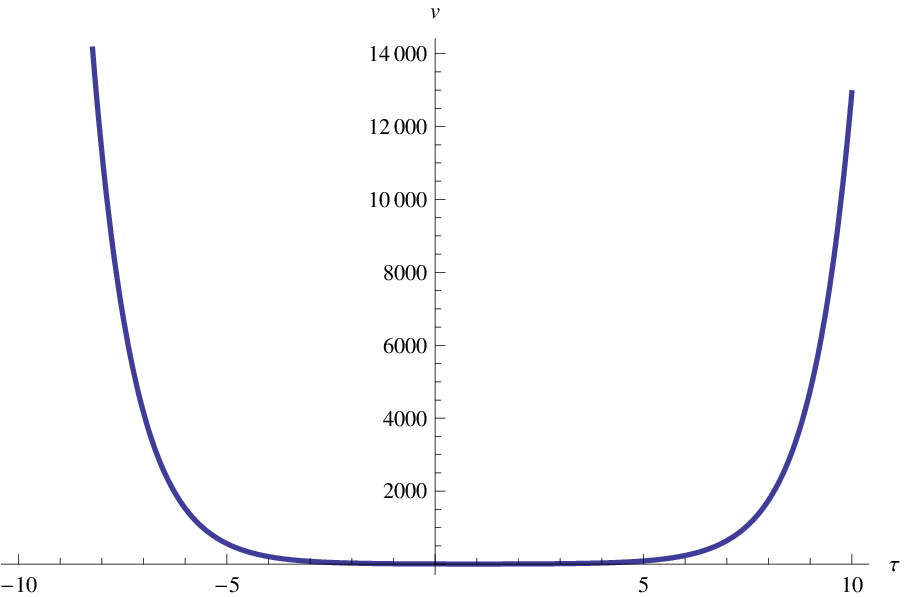}
\hspace{5mm}
\includegraphics[height=35mm]{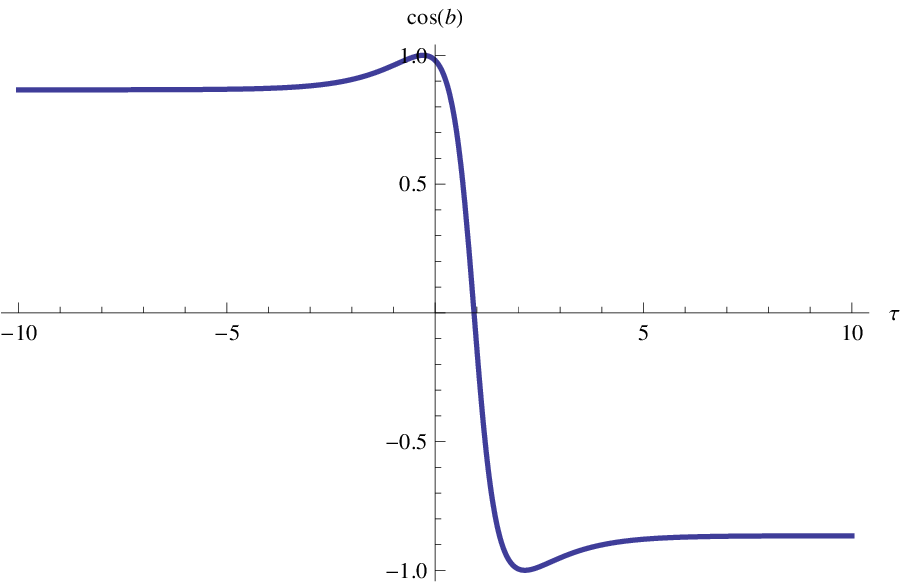}
\hspace{5mm}
\includegraphics[height=35mm]{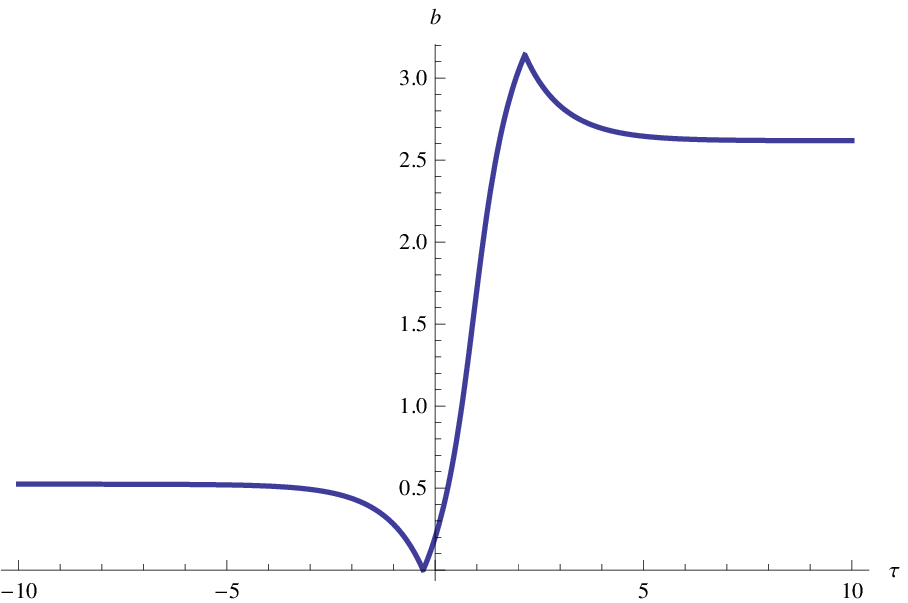}
\caption{Plots of the volume $v$ (on the left), of $\cos b$ (in the center) and of the conjugate momentum $b$ (on the right) evolving as functions of the internal time $\tau$, for $\tau_0=0$, for explicit values of the parameters: $\beta=0.5$, $v_m=1$, $v_0=3$, $b_0=0.2$, $H_{reg}\sim 2.062$, $A\sim4.375$ and $B\sim-3.201$.
\label{hyper_plot}}
\end{center}
\end{figure}

\item $\beta>1$: time-like Hamiltonian

In this case the evolution is given by the following rotation
\be
U_\tau=\left( \begin{array}{cc}
\cos\f\ttau2+\f{i\beta}{\sqrt{\beta^2-1}}\sin\f\ttau2 & \f{1}{\sqrt{\beta^2-1}}\sin\f\ttau2\vspace*{0.2cm} \\
\f{1}{\sqrt{\beta^2-1}}\sin\f\ttau2 &
\cos\f\ttau2-\f{i\beta}{\sqrt{\beta^2-1}}\sin\f\ttau2\end{array} \right),
\ee
with $\ttau=\tau\sqrt{\beta^2-1}$.
The resulting trajectories are similar to the previous case but replacing the hyperbolic functions by trigonometric functions (see plots below in fig. \ref{trigo_plot}):
\be
v(\tau)
\,=\,
A\cos(\ttau-\ttau_0)+B\sin(\ttau-\ttau_0)-\f{\beta}{1-\beta^2}H_{reg}\,,
\ee
\be
\cos b(\tau)
\,=\,
\f{\sqrt{\beta^2-1}}{\sqrt{v(\tau)^2-v_m^2}}\,
\left[A\sin(\ttau-\ttau_0)-B\cos(\ttau-\ttau_0)
\right]\,,
\ee
where the constants $A$ and $B$ are determined in terms of the initial conditions $v_0$ and $b_0$, or equivalently in terms of $v_0$ and the energy $H_{reg}$.
%

In this case, the volume is completely bounded, the universe oscillates between a minimal volume (bounce) and a maximal volume and its motion is periodic.

\begin{figure}[h]
\begin{center}
\includegraphics[height=35mm]{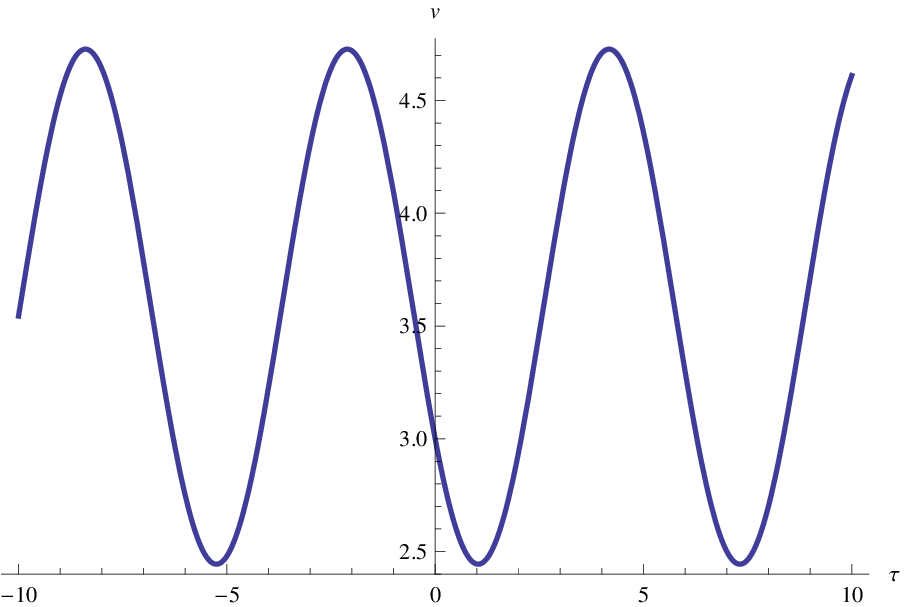}
\hspace{5mm}
\includegraphics[height=35mm]{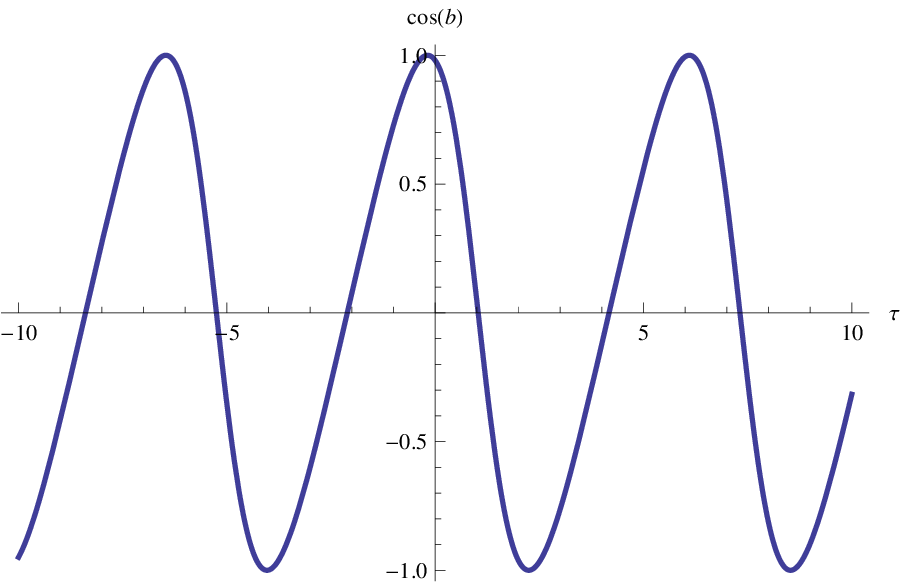}
\hspace{5mm}
\includegraphics[height=35mm]{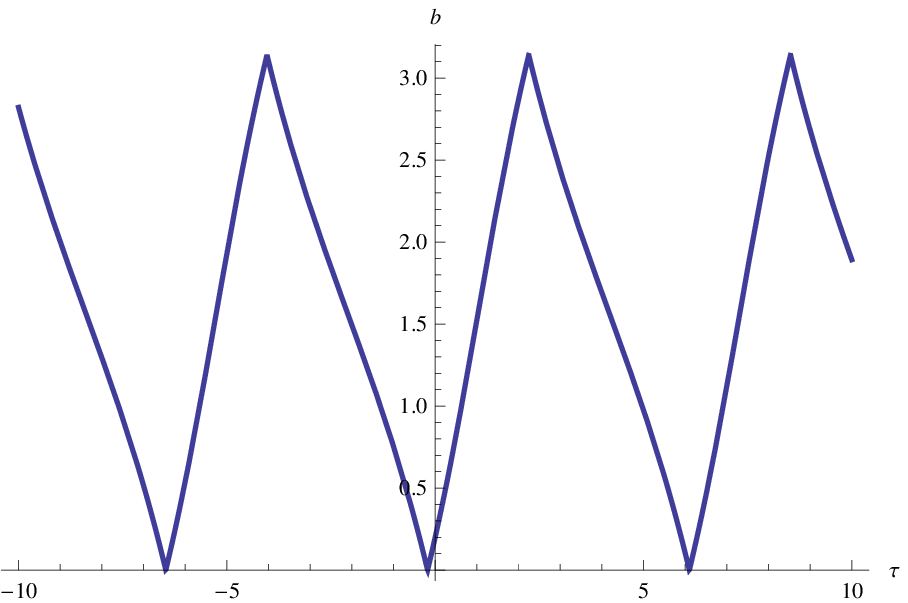}
\caption{Plots of the volume $v$ (on the left), of $\cos b$ (in the center) and of the conjugate momentum $b$ (on the right) evolving as functions of the internal time $\tau$, for $\tau_0=0$, for explicit values of the parameters: $\beta=3$, $v_m=1$, $v_0=3$, $b_0=0.2$, $H_{reg}\sim 9.562$, $A\sim-0.586$ and $B\sim-0.980$.
\label{trigo_plot}}
\end{center}
\end{figure}

\end{itemize}

If one compares these trajectories with the ones of classical flat FRW cosmology with non-vanishing cosmological constant, then qualitatively it seems that the case $|\beta|<1$ corresponds to $\Lambda>0$ while the case $|\beta|>1$ corresponds to a negative cosmological constant $\Lambda<0$. Indeed, as described in appendix \ref{C2}, for $\Lambda>0$, classical FRW cosmology gives two possible branches: either a universe contracting from infinite volume and crashing to zero volume, or a universe born in a big bang and expanding to infinite volume. Here our regularized dynamics has a big bounce which connects the contracting phase to the expanding phase, without going through a singularity. On the other hand, for $\Lambda<0$, classical FRW cosmology describes a universe expanding from a big bang, then reaching a maximal volume before crashing in a big crunch. Our regularized dynamics once again avoids the vanishing volume singularity and creates cycles oscillating between minimal volume and maximal volume.

However, if one now compares the explicit equations of the trajectories, as given above to the ones for classical FRW cosmology given in appendix \ref{C2}, one sees that the formulas do not look the same at all. Indeed, quantitatively, it is not clear that there is a precise regime where our regularized $\SU(1,1)$ dynamics matches the classical FRW models.

We see a few possible reasons for this mismatch and propose potential ways to remedy them:
\begin{itemize}
\item The $\su(1,1)$ Hamiltonian is just not enough. As we have seen the Friedmann equations (in internal time) are just different and they do not seem to match for large volume. One should probably depart from the strict $\SU(1,1)$ evolution. In particular, we should investigate the exact ansatz for the effective dynamics for loop quantum cosmology with cosmological constant and see how to take it into account in our framework.

\item The choice of the internal time is not the correct one to compare the classical dynamics to our regularized model. For instance, the classical FRW cosmology in proper time seems to match the regularized trajectories in internal time. But we do not yet see a clear mathematical reason for this nor a physical motivation for this switch of cosmological clock.

\item It is a problem of domain of validity of the regime in which we compared our Hamiltonian initially: we were looking at first at small fluctuations in $b$ around some fixed value $b_0$. After solving explicitly the equations of motion, we see that $b$ varies too much generically in (internal) time and thus causes some large deviation of the regularized dynamics from the classical trajectories. Nevertheless, one could try to identify a precise regime (possibly looking at the evolution in terms of a different clock) for intermediate values of the volume where $b$ does not fluctuate wildly and where the trajectories would match analytically.

\item Finally, maybe the $J_z$ term simply does not correspond to the inclusion of a cosmological constant but of another gravitational source (matter field or dark energy or...). This would require studying the coupling of various matter fields to the effective dynamics of loop cosmology.

\end{itemize}

We postpone a detailed analysis of these alternatives to future investigation. Our purpose here is to focus on the $\SU(1,1)$ structure of the effective dynamics for loop cosmology and to see how far one can get with it. Since it seems likely that one has to add interactions which lead to deviation from the exact $\SU(1,1)$ flow, it seems more reasonable to leave this for later.

\subsection{Generalizing the $K_y$ Hamiltonian: Accounting for Curvature?}

We have investigated above the addition of $K_x$ or $J_z$ terms. We have found the $K_x$ term physically irrelevant and the $J_z$ term potentially related to the inclusion of a cosmological constant. However, one can naturally wonder if it is possible to account for non-flat FRW cosmology with $k=\pm 1$.

An interesting approach is the possibility of generalizing further our description without departing from the $H=K_y$ Hamiltonian. Indeed, it is easy to realize that the set of observables \eqref{obs} is not the most general one forming a closed $\su(1,1)$ algebra, but this set can be generalized to
\begin{align}
 J_z=v, \qquad K_+=\sqrt{v^2-v_m^2} e^{i(b+\psi(v))},\qquad K_-=\sqrt{v^2-v_m^2} e^{-i(b+\psi(v))},
\end{align}
for any real function $\psi(v)$. It might happen that for a suitable choice of $\psi(v)$, the Hamiltonian
\be
H_{reg}=K_y=\sqrt{v^2-v_m^2} \sin(b+\psi(v))
\ee
provides a regularized dynamics for some FRW model. Actually, for the particular choice, inspired from the models of effective dynamics for loop quantum cosmology,
\be
\psi(v)=-\f{k}{\gamma(4\pi G v)^{1/3}}
\ee
and in the regime of small $b$ and large $v$ the Hamiltonian behaves as
\be
H_{reg}\sim vb-\f{k}{\gamma(4\pi G)^{1/3}}v^{2/3},
\ee
which corresponds to a Hamiltonian constraint
\be
p_\phi^2\sim 12\pi G \left[b^2v^2-\f{2k}{\gamma}\f{bv^{5/3}}{(4\pi G)^{1/3}}+\f{k^2}{\gamma^2(4\pi G)^{2/3}}v^{4/3}\right],
\ee
for the case that the geometry is coupled to a massless scalar field and $H_{reg}=p_\phi/\sqrt{12\pi G}$.
The above expression almost coincides with the Hamiltonian constraint of the FRW model with curvature $k$ and without cosmological constant, as we can see in \eqref{const}. Therefore, it is natural to analyze whether indeed $H_{reg}$ provides a regularized non-flat FRW model.

Using \eqref{evol} it is straightforward to get the trajectories, they are given by
\begin{align}
v(\tau)=v(\tau_o)\cosh(\tau-\tau_o)\,,\qquad
b(\tau)=\arccos\frac{v(\tau_o)\sinh(\tau-\tau_o)}{\sqrt{v^2(\tau)-v_m^2}}+\f{k}{\gamma[4\pi G v(\tau)]^{1/3}}\,
\end{align}
Qualitatively, these trajectories seem to provide a regularized version of the FRW model with non-zero curvature, avoiding the vanishing volume singularity. On the other hand, the actual explicit behavior of $v$ and $b$ in terms of $\tau$ in our regularized version does not compare at all with the exact classical FRW trajectories as computed in appendix \ref{C3}. We do not fully understand this mismatch and how to exactly resolve this issue. The various alternatives that we see are the same as given above in the case of the $J_z$ term and the cosmological constant.

\section*{Conclusion}

In the present work we have carried out a group theoretical quantization of the flat FRW model coupled to a massless scalar field adopting the regularizations employed in the improved dynamics of LQC, both in the kinematical volume and in the variable conjugate to it. This group theoretical quantization lies in the fact that the set of observables that describes the regularized phase space close an $\su(1,1)$ algebra. The preservation of the $\su(1,1)$ structure at the quantum level provides a quantum representation of the algebra of classical observables free of anomalies and free of factor ordering ambiguities. In particular, it fixes totally the Hamiltonian constraint operator. The irreducible representations of the group $\SU(1,1)$ of the discrete principal series provide superselection sectors. In each sector we have explicitly constructed dynamical coherent states to analyze the evolution. We have shown that, in these coherent states, the volume undergoes a bounce that cures the classical big bang singularity, and that the relative fluctuations of the volume remain bounded along the whole evolution.

Furthermore, we have investigated whether our $\su(1,1)$ framework can be generalized to account for the introduction of cosmological constant or curvature. Our analysis shows that the models with curvature or with cosmological constant are more complicated, and indeed a quantization of them within the pure $\su(1,1)$ structure does not seem plausible. In order to get a group quantization for those more general models we would need to depart from the $\su(1,1)$ algebra by considering its enveloping algebra.

Another intriguing feature is the generalization of the $\su(1,1)$ structure of the algebra of observables to more complicated algebras. For instance, beyond the $\su(1,1)$ algebra, one can consider all the observables $L_n=v e^{inb}$ for $n\in \Z$, which obviously form a Witt algebra. One can wonder if this allows to take in account and explicitly solve a larger class of cosmological Hamiltonians, and whether a central extension to a Virasoro algebra would have any physical meaning. Finally, it would be interesting to see if our group theoretical approach to the loop quantization of cosmological models can be pushed further and whether it is possible to identify relevant Lie algebra structures in the space of observables for Bianchi models \cite{bianchi} or Gowdy cosmologies \cite{gowdy1,gowdy2}.

\section*{Acknowledgments}

We would like to thank Victor Aldaya, Martin Bojowald, Guillermo Mena Marug\'an and Edward Wilson-Ewing for their constructive comments and their encouragements.

EL is partially supported by the ANR ``Programme Blanc" grant LQG-09. MMB is partially supported by the Spanish MICINN Project No. FIS2011-30145-C03-02.

\appendix

\section{Spinorial Representation for $\SU(1,1)$}
\label{AppA}
\subsection{The 2-Dimensional Representation and $\SU(1,1)$ matrices}
\label{A1}
The Lie group $\SU(1,1)$ is defined as the set of 2$\times$2 matrices $U$ of determinant $\det U=1$ satisfying $U\eps U^\dagger = \eps$ with:
$$
\eps=\mat{cc}{ 1 & 0 \\ 0 & -1 }\,.
$$
Explicitly, the group elements read:
\be
U=\mat{cc}{\alpha & \beta \\ \bar{\beta} & \bar{\alpha}},
\qquad\textrm{with}\quad
|\alpha|^2-|\beta|^2=1\,.
\ee
The action of such matrices on complex vector $(x,y)\in\C^2$ conserves the pseudo-norm $|x|^2-|y|^2$:
\be
\mat{c}{x\\ y}
\,\arr\,
U\mat{c}{x\\ y}=\mat{cc}{\alpha & \beta \\ \bar{\beta} & \bar{\alpha}}\mat{c}{x\\ y},
\qquad
|x|^2-|y|^2
=\mat{c}{x\\ y}\dag\eps \mat{c}{x\\ y}
=\mat{c}{x\\ y}\dag U\dag\eps U\mat{c}{x\\ y} \,.
\ee
The generators of $\su(1,1)$ are the (Lorentzian) Pauli matrices:
\be
\label{Pauli}
\sigma_z=\f12\mat{cc}{1 & 0 \\ 0 & -1},\quad
\sigma_x=\f12\mat{cc}{ 0 & 1 \\ -1 & 0 },\quad
\sigma_y=\f{-i}2\mat{cc}{0 & 1 \\ 1 & 0 }\,.
\ee
Their commutators define  the $\su(1,1)$  Lie algebra:
$$
[\sigma_z,\sigma_x]=+i\sigma_y,\quad
[\sigma_z,\sigma_y]=-i\sigma_x,\quad
[\sigma_z,\sigma_x]=-i\sigma_z\,.
$$
Then group elements are obtained  through exponentiation, $U=\exp\,i\vu\cdot\vsigma$. More explicitly, we encounter three cases. If the 3-vector $\vu$ is null, $|u|^2=u_z^2-u_x^2-u_y^2=0$, then the matrix $(\vu\cdot\vsigma)$ is nilpotent and the series expansion is truncated at first order:
\be
|u|^2=0\,\Rightarrow\quad
e^{i\vu\cdot\vsigma}=\id+i\vu\cdot\vsigma.
\ee
If the 3-vector $\vu$ is time-like, we get a trigonometric expression:
\be
|u|^2>0\,\Rightarrow\quad
e^{i\vu\cdot\vsigma}=
\cos\f{|u|}2\id+2i\sin\f{|u|}2\,\f{\vu}{|u|}\cdot\vsigma.
\ee
For space-like vectors, we get a similar expression but with hyperbolic functions.

\subsection{Phase Space Representation of the $\su(1,1)$ Lie Algebra}

\label{A2}

Let us start with the four dimensional phase space defined by two complex variables $z^{0,1}\in\C$ and equipped with the canonical Poisson bracket:
\be
\{z^0,\bz^0\}=\{z^1,\bz^1\}=-i.
\ee
Then we consider the following observables:
\be\label{obsz}
J_z=\f12\,\left(|z^0|^2+|z^1|^2\right),
\qquad
K_+=\bz^0\bz^1,
\qquad
K_-=z^0z^1,
\qquad
L=\f12\,\left(|z^0|^2-|z^1|^2\right)\,.
\ee
It is easy to check that the three observables $J_z,K_\pm$ form a $\su(1,1)\sim\sl_2$ Lie algebra while $L$ commutes with all three of them:
\be
\{J_z,K_\pm\}=\,\mp i K_\pm,\qquad
\{K_+,K_-\}=\,2iJ\,,
\ee
\be
\{L,J_z\}=\{L,K_\pm\}=0\,.
\ee
Instead of $K_\pm$, one can use the usual generator of boosts in the $x$ and $y$ directions:
\be\label{realobsz}
K_\pm=K_x\pm i K_y,\qquad
K_x=\f12(K_++K_-),\qquad
K_y=\f1{2i}(K_+-K_-),
\ee
which satisfy the following commutation relations:
$$
\{J_z,K_x\}=K_y,
\qquad
\{J_z,K_y\}=-K_x,
\qquad
\{K_x,K_y\}=-J_z\,.
$$

Noting $\vJ=(K_x,K_y,J_z)\in\R^{2,1}$ for the 3-vector living in the three dimensional Minkowski space of signature $(--+)$, its norm defines the Casimir of the $\su(1,1)$ algebra and is simply expressed in terms of the $L$-observable:
\be\label{L}
C\,\equiv\, J^2-K_x^2-K_y^2\,=\,J_z^2-K_+K_- \,=\, L^2\,,
\ee
so that we only generate time-like or null vectors with $C=\vJ^2\ge 0$. Null vectors $\vJ^2=0$ correspond to complex variables with equal norm, $|z^0|=|z^1|$.

\subsection{Action of $\SU(1,1)$ Transformations}
\label{SU11spinor}

In order to derive the action of finite $\SU(1,1)$ transformations on our variables, let us start by looking at the action of the $\su(1,1)$ generators on $z^{0,1}$ and $\bz^{0,1}$. They mix the variables and their complex conjugate. Nevertheless, we easily notice that they mix $z^0$ only with $\bz^1$ and $z^1$ only with $\bz^0$. It thus seems natural to introduce the following spinor $z\in\C^2$:
\be
|z\ra=\mat{c}{z^0\\ \bz^1},\qquad
\la z| =\mat{cc}{\bz^0 & z^1}.
\ee
It is direct to compute the action of the generators on this complex 2-vector:
\be
\{\vJ,|z\ra\}
\,=\,
i\,\vsigma\,|z\ra,
\ee
where $\vsigma$ are the Lorentzian Pauli matrices defined in \eqref{Pauli}.
It is then straightforward to exponentiate the action of the $\vJ$'s and check that the spinor $z$ does belong to this fundamental two-dimensional representation of  $\SU(1,1)$:
\be
e^{\{\vu\cdot\vJ,\,.\,\}}\,|z\ra
\,=\,
U\,|z\ra,\qquad
U=e^{i\vu\cdot\vsigma}\in\SU(1,1)\,.
\label{SU11action}
\ee
If we switch the role of $z^0$ and $z^1$ and now take the complex conjugate $\bz^0$  in the definition of the spinor, it simply amounts to taking the complex conjugate of the spinor $|z\ra$ and it transforms as in the complex conjugate representation:
\be
\overline{|z\ra}=|\bz\ra=\mat{c}{\bz^0\\ z^1}
\quad
\longrightarrow
\quad
e^{\{\vu\cdot\vJ,\,.\,\}}\,|\bz\ra
\,=\,
\bar{U}\,|\bz\ra\,.
\ee

\medskip

From these finite $\SU(1,1)$ transformation, one can check directly that $L(z)=\f12(|z^0|^2-|z^1|^2)$ is indeed conserved:
$$
U=\mat{cc}{\alpha & \beta \\ \bar{\beta} & \bar{\alpha}}\,\in\SU(1,1),\qquad
L(Uz)=L(z)\,,
$$
reflecting the fact that the observable $L$  commutes with the $\su(1,1)$ generators.
\medskip

Next, we introduce the $2\times 2$ matrix:
\be
M\,\equiv\,
\mat{cc}{J_z & K_- \\ K_+ & J_z}\,.
\ee
It admits a simple expression in terms of the spinor $z$:
\be
M\,=\,
|z\ra\la z| - L(z)\,\eps
\,.
\ee
From the law of transformation of the spinor and  the facts that $L(z)$ is invariant under $\SU(1,1)$ and that $U\eps U^\dagger = \eps$ for any matrix in $\SU(1,1)$ by definition, one find that $M$ lives in the adjoint representation~\footnotemark:
\be
e^{\{\vu\cdot\vJ,\,.\,\}}\,M
\,=\,
U M U^\dagger\,,
\quad
U=e^{i\vu\cdot\vsigma}\,.
\ee
\footnotetext{
One could have check this by directly computing the Poisson bracket of $M$ with the $\su(1,1)$ generators:
$$
\{\vJ,M\}
\,=\,
i(\vsigma M - M \vsigma^\dagger)\,.
$$}
%

\section{Time-like Representations of  $\SU(1,1)$ and Coherent States}
\label{AppB}

\subsection{Deriving Unitary Representations From Harmonic Oscillators}
\label{B1}

Let us quantize the phase space defined above and thus promote the complex variables $z^{0,1}$ and their complex conjugate $\bz^{0,1}$ to respectively annihilation operators $a,b$ and their corresponding creation operators $a\dag,b\dag$, satisfying the canonical commutation relations:
$$
[a,a\dag]=[b,b\dag]=1,\quad
[a,b]=0\,.
$$
Following this simple quantization rule, we define the $\su(1,1)$ generators:
\be
J_z=\f12\left(a\dag a+b\dag b +1\right),
\qquad
K_+=a\dag b\dag,
\qquad
K_-=ab\,.
\ee
The $+1$ term in $J_z$ comes from properly ordering the operators, and one checks that these form a $\su(1,1)$ Lie algebra:
$$
[J_z,K_\pm]=\pm K_\pm,\qquad
[K_+,K_-]=-2J\,.
$$
The Casimir of the algebra admits a simple expression in terms of the quantized version of $L$:
\be
C=J^2-\f12\left(K_+K_-+K_-K_+\right)
=L^2-\f14=\left(L+\f12\right)\left(L-\f12\right),
\qquad
L=\f12\left(a\dag a-b\dag b \right)\,.
\ee
Then irreducible representations of $\SU(1,1)$ will be determined by the value of the operator $L$, which measures the (fixed) difference of energy between the two harmonic oscillators.

\medskip

Working with the two quantum oscillators, our Hilbert space is the tensor product of the two Hilbert spaces for the decoupled oscillators, $\cH_{HO}\otimes\cH_{HO}$. Working with the standard basis diagonalizing the number of quanta $a\dag a$ and $b\dag b$ of both oscillators, we can compute the action of the operators $\vJ$ and $L$:
\beq
L|n_a,n_b\ra_{HO}
&=&
\f12(n_a-n_b) |n_a,n_b\ra_{HO},
\nn\\
J_z|n_a,n_b\ra_{HO}
&=&
\f12(n_a+n_b+1) |n_a,n_b\ra_{HO},
\nn\\
K_+|n_a,n_b\ra_{HO}
&=&
\sqrt{(n_a+1)(n_b+1)} |n_a+1,n_b+1\ra_{HO},
\nn\\
K_-|n_a,n_b\ra_{HO}
&=&
\sqrt{n_an_b} |n_a-1,n_b-1\ra_{HO}\,.\nn
\eeq
In order to get an irreducible representation of $\SU(1,1)$, we diagonalize the operator $L$. This fixes the difference of energy between the two oscillators, let us say to $N=n_a-n_b$. Let us start with $N\ge 0$. The corresponding Casimir is $C=(N-1)(N+1)/4=j(j-1)$ with the spin $j\equiv (N+1)/2$ always larger or equal to $\f12$. The usual $\SU(1,1)$  basis is defined by diagonalizing the operator $J_z$. Its eigenvalue defines the magnetic momentum $m=(n_a+n_b+1)/2$ always equal or larger than the spin $j$:
\be
|j,m\ra=|n+N,n\ra_{HO},
\qquad\textrm{with}\quad
N=2j-1,\quad j=\f {(N+1)}2\ge\f12,\quad
n=m-j,\quad m=j+n\ge j\,.
\ee
Thus the irreducible representation of $\su(1,1)$ of spin $j$ lives on the Hilbert space spanned by the basis states $|j,m\ra$ with $m\ge j$ bounded from below, $\cV_+^j\equiv\bigoplus_{m\ge j} \C\,|j,m\ra$.
Then it is straightforward to compute the action of the $\su(1,1)$ generators:
\beq
J_z|j,m\ra
&=&
m|j,m\ra\,,\\
K_+|j,m\ra
&=&
\sqrt{(m-j+1)(m+j)}|j,m+1\ra\,,\nn\\
K_-|j,m\ra
&=&
\sqrt{(m-j)(m+j-1)}|j,m-1\ra\,.\nn
\eeq

The representations with negative value of $L$ have the $|n,n+N\ra$ with fixed $N\ge 0$ and $n\in\N$ as basis states. They lead to isomorphic irreducible representations.

In order to get the highest weight representations $\cV_-^j\equiv\bigoplus_{m\le -j} \C\,|j,m\ra=\overline{\cV_+^j}$, one needs to re-define the $\su(1,1)$ generators in terms of the harmonic oscillator operators. Indeed re-defining the generators as $J_z \arr -J_z$ and $K_\pm\arr K_\mp$, they still satisfy the same commutation relations, but we now get negative eigenvalues for $J_z$ and obtain the dual representation.

\medskip

Using this method, we generate all time-like unitary irreducible representations of $\SU(1,1)$, with $C+\f14\ge 0$. It does not however allow us to generate the space-like unitary representation with $C+\f14<0$. Anyhow, only the time-like representations of $\SU(1,1)$ are involved in the quantization of the effective/regularized LQC dynamics for FRW cosmology.

\subsection{Defining Coherent States}
\label{B2}

Let us define the following states living in the $\cV_+^j$ representation as defined above and labeled by the classical spinor $z\in\C^2$:
\be
|j,z\ra\,\equiv\,
\sum_{n=0}^\infty \sqrt{\f{(n+N)!}{n!N!}}\,
\left(\f1{\bz^0}\right)^{n+N+1}\,(z^1)^n\,|n+N,n\ra_{HO}\,.
\ee
As we show below, these states are coherent in that they transform covariantly under $\SU(1,1)$ transformations and they are semi-classical states peaked on classical phase space points with minimal uncertainty.

For the special case, where the spinor is trivial, $z^0=1$ and $z^1=0$, the state reduces to the lowest weight vector:
\be
|j,(z^0,z^1)=(1,0)\ra
\,=\,
|N,0\ra_{HO}
\,=\,
|j,m=j\ra\,.
\ee

In order to compute the norm and expectation values of those states, we simply need the following Taylor series:
\be
\sum_n \f{(n+N)!}{n!N!}\,X^n
\,=\,
\f1{(1-X)^{N+1}}\,.
\ee
This allows to compute the norm (the series converge when $|z^0|>|z^1|$):
\be
\la j,z|j,z\ra
=\f1{|z^0|^{2(N+1)}}\sum_n\f{(n+N)!}{n!N!} \left(\f{|z^1|^2}{|z^0|^2}\right)^n
=\f1{(|z^0|^2-|z^1|^2)^{N+1}}
=\f1{(2L(z))^{2j}}\,,
\ee
which is invariant under $\SU(1,1)$ transformations. Then we compute the expectation values of the $\su(1,1)$ generators:
\be\label{J}
\la j,z|J_z|j,z\ra
=\f1{|z^0|^{2(N+1)}}\sum_n\f{(n+N)!}{n!N!} \,(n+j)\,\left(\f{|z^1|^2}{|z^0|^2}\right)^n
=\la j,z|j,z\ra\,j\,\f{|z^0|^2+|z^1|^2}{|z^0|^2-|z^1|^2}
\quad
\Rightarrow
\,\,
\la J_z\ra=
j\,\f{J_z}{L}\,,
\ee
\be
\la j,z|K_+|j,z\ra
=\f{\bz^0}{z^1}\f1{|z^0|^{2(N+1)}}\sum_n\f{(n+N)!}{n!N!} \,n\,\left(\f{|z^1|^2}{|z^0|^2}\right)^n
=\la j,z|j,z\ra\,\f {2j\bz^0\,\bz^1}{|z^0|^2-|z^1|^2}
\quad
\Rightarrow
\,\,
\la K_+\ra=
j\,\f{K_+}{L}\,,
\ee
\be
\la j,z|K_-|j,z\ra
=\f{z^1}{\bz^0}\f1{|z^0|^{2(N+1)}}\sum_n\f{(n+N)!}{n!N!} \,(n+2j)\,\left(\f{|z^1|^2}{|z^0|^2}\right)^n
=\la j,z|j,z\ra\,\f {2jz^0\,z^1}{|z^0|^2-|z^1|^2}
\quad
\Rightarrow
\,\,
\la K_-\ra=
j\,\f{K_-}{L}\,,
\ee
Thus one gets exactly the expected classical vector up to a simple global re-scaling:
\be
\la \vJ \ra= j\,\f{\vJ}{L},
\ee
so that the norm of  $\la \vJ \ra$ only depends on the spin $j$ of the chosen representation:
\be
\la \vJ \ra^2=j^2\,.
\ee
Since $\vJ^2$ is the Casimir and its value is already known, this allows to compute the invariant fluctuation of our coherent states:
\be
\la \vJ^2 \ra-\la \vJ \ra^2= j(j-1)-j^2=-j,
\ee
which is actually the minimal possible fluctuation for a time-like representation\footnotemark.
\footnotetext{
A rough calculation on the standard basis states gives:
$$
\la j,m|\vJ^2|j,m\ra-\la j,m|\vJ|j,m\ra^2=
j(j-1)-m^2,
$$
which is obviously minimal for the lowest weight vector $m=j$.
}
We can further compute the fluctuations for the individual components. We get:
\be\label{DJ}
\la J_z^2\ra-\la J_z\ra^2
\,=\,
2j\,\f{|z^0|^2|z^1|^2}{(|z^0|^2-|z^1|^2)^2}
\,=\,
\f j2\,\f{K_+K_-}{L^2}
\,=\,
\f j2\,\left(\f{J_z^2}{L^2}-1\right)\,,
\ee
\be\label{DKx}
\la K_x^2\ra-\la K_x\ra^2
\,=\,
\f j2\,\f{(|z^0|^2-|z^1|^2)^2+(\bar{z}^0\bar{z}^1+z^0z^1)^2}{(|z^0|^2-|z^1|^2)^2}
\,=\,
\f j2\,\left(\f{K_x^2}{L^2}+1\right)\,,
\ee
\be\label{DKy}
\la K_y^2\ra-\la K_y\ra^2
\,=\,
\f j2\,\f{(|z^0|^2-|z^1|^2)^2-(\bar{z}^0\bar{z}^1-z^0z^1)^2}{(|z^0|^2-|z^1|^2)^2}
\,=\,
\f j2\,\left(\f{K_y^2}{L^2}+1\right)\,.
\ee

Furthermore, these coherent states saturate the uncertainty relations. Let us remind that given two self-adjoint operators $A$ and $B$ they satisfy the following uncertainty relation:
\be
(\la A^2 \ra-\la A \ra^2)(\la B^2 \ra-\la B \ra^2)+\f1{4}\la [A,B] \ra^2\geq \left(\f12\la AB+BA\ra-\la A \ra\la B \ra\right)^2\,.
\ee
After a tedious but straightforward calculation one can indeed check that the above relation, when particularized to any two of the three operators $J_z$, $K_x$ and $K_y$, becomes an identity.
Thus our $\SU(1,1)$ coherent states saturates all uncertainty relations for the $\su(1,1)$ generators.

\subsection{Resolution of the Identity}

\label{B3}

These coherent states provide a decomposition of the identity on the Hilbert space $\cV_+^j$., for each fixed value of $L$. Indeed, let fix $L=l$, then the two complex variables are related to one another by $|z^0|^2=2l+|z^1|^2$. Then we compute the following integral:
\beq
\int {d^2z^0d^2z^1}\,\delta(L-l)\,
|j,z\ra\la j,z|
&=&
\int d^2z^0d^2z^1
\sum_{n,\tn}
\sqrt{\f{(n+N)!}{n!N!}}\,\sqrt{\f{(\tn+N)!}{\tn!N!}}\,
\f{2\delta(2L(z)-2l)\,(z^1)^n(\bz^1)^\tn}{(\bz^0)^{n+N+1}(z^0)^{\tn+N+1}}\,|n+N,n\ra\la\tn+N,\tn|\nn\\
&=&
\sum_{n}
\f{(n+N)!}{n!N!}\,
\int d^2z^0d^2z^1\,2\delta(|z^0|^2-|z^1|^2-2l)
\f{|z^1|^{2n}}{|\bz^0|^{2(n+N+1)}}\,|n+N,n\ra\la n+N,n|\nn\\
&=&
(2\pi)^2\sum_{n}
\f{(n+N)!}{n!N!}\,
\int rdr\,
\f{r^{2n}}{(2l+r^2)^{n+N+1}}\,|n+N,n\ra\la n+N,n|\nn\\
&=&
\f{(2\pi)^2l}{N(2l)^{N+1}}
\,\sum_{n}|n+N,n\ra\la n+N,n|
\eeq

Up to the pre-factor, which only depends on the choice of representation (through $N$) and the specific value of $L$, we do have in the end a proper decomposition of the identity on $\cV^j_+$.

\subsection{Action of $\SU(1,1)$ on the Coherent States}

The key property of these coherent states is that they transform covariantly under $\SU(1,1)$ transformations and that their shape remains undeformed under the action of $\SU(1,1)$. More explicitly, we have:
\be
U\,|j,z\ra=|j,U\vartriangleright z\ra,\qquad U=e^{i\vu\cdot\vJ}\,\in\SU(1,1),
\ee
for arbitrary $\SU(1,1)$ transformations where $U\vartriangleright z$ is the action \eqref{SU11action} defined above in section \ref{SU11spinor}.
It is straightforward to check this property for infinitesimal transformation around the identity, $U\sim\id$, then one can exponentiate that action.

This property ensures that all the coherent states $|j,z\ra$ with $2L(z)=1$ are obtained from the lowest weight vector $|j,j\ra$ by a $\SU(1,1)$ transformation:
\be
|j,j\ra=|j,(z^0,z^1)=(1,0)\ra,
\qquad
|j,z\ra=U(z)\,|j,j\ra,
\ee
$$
\textrm{with}
\quad
\mat{c}{z^0\\\bz^1}
=U(z)\,\mat{c}{1\\0}
=\mat{cc}{\alpha&\beta\\\bbeta &\balpha}\mat{c}{1\\0}
=\mat{c}{\alpha\\ \bbeta},
\quad
2L=|z^0|^2-|z^1|^2=|\alpha|^2-|\beta|^2=1,
\quad
U(z)\in\SU(1,1)\,.
$$
To obtain coherent states with $2L\ne1$, one need to check the scaling properties of $L$ and the coherent states under the re-scaling transformation $z\arr\lambda z$ with $\lambda>0$. For instance, we have, $L(\lambda z)=\lambda^2L(z)$ and:
$$
|j,\lambda z\ra=\lambda^{-2j}\,|j, z\ra\,.
$$
Thus to get an arbitrary coherent state $|j,z\ra$ from $|j,j\ra$, one simply has to do a re-scaling and a $\SU(1,1)$ transformation (remember that our coherent states are well-defined only for $|z^0|^2> |z^1|^2$ i.e $L(z)> 0$):
\be
|j,z\ra
\,=\,
(\sqrt{2L})^{-2j}\,U\left(\f z{\sqrt{2L}}\right)\,|j,j\ra.
\ee

The fact that these states are all obtained from $|j,j\ra$ through straightforward $\SU(1,1)$ transformations (up to an over-all factor) naturally implies that their invariant uncertainty $\la \vJ^2\ra -\la\vJ\ra^2$ as computed above is equal to the uncertainty associated to the state $|j,j\ra$, that is $-j$.

\subsection{Gaussian Approximation for the Coherent States}

For a fixed spinor $z$, let us look on the coefficients $\la j,m|j,z\ra$ in terms of $m$. We will see that for appropriate spinors, this distribution can be approximated as a phased Gaussian, making it similar to the standard ansatz for coherent states.

We fix the representation $N$ and use the Stirling formula for large $n$'s:
\beq
\la n+N,n|j,z\ra
&=&
\sqrt{\f{(n+N)!}{n!N!}}\f1{(\bz^0)^{n+N+1}}(z^1)^n\nn\\
&\sim&
\f1{(\bz^0)^{N+1}\sqrt{N!}}
\left(1+\f N n\right)^{\f14}n^{\f N2}\left(\f{z^1}{\bz^0}\right)^n\nn\\
&\sim&
\f1{(\bz^0)^{N+1}\sqrt{N!}}
e^{\f N2\log n - n\log \xi},
\quad
\textrm{with}
\quad
\xi=\f{\bz^0}{z^1}\,.\nn
\eeq
The exponent $\phi(n)\,\equiv\,\f N2\log n - n\log \xi$ has a unique (complex) extrema:
$$
\pp_n\phi=\f N{2n}-\log\xi
\quad\Rightarrow\quad
n_0=\f N{2\log\xi}.
$$
At this point, it is convenient to use radial coordinates for the complex number $\xi$:
$$
\xi=R e^{i\psi}=\f{\sqrt{r^2+2L}}{r}e^{i\psi},\qquad
\log\xi\equiv \log R +i\psi,
$$
where $r$ and $\sqrt{r^2+2L}$ are respectively the modulus of $z^1$ and $z^0$. Now let us remember that our coherent states are well-defined for $L(z)\ge 0$, thus for $R>1$. Then for large values of $r$, i.e small values of $R\arr 1_+$, the logarithm $\log R$ becomes small and the extremal value of $n$ grows inversely to $\log R$ and thus becomes large, justifying the Stirling approximation for the factorials.

Computing the value of the second derivative $\pp^2\phi$ at the extremum, we can finally give the stationary point approximation for our distribution:
\be
\la n+N,n|j,z\ra
\,\underset{n\gg 1}{\sim}\,
e^{\f N2\left(\log\f{N}{2\log\xi}-1\right)}\,
e^{-\f{(\log \xi)^2}{N}\,\left(n-\f{N}{2\log\xi}\right)^2},
\ee
which is a phased Gaussian peaked on the real value $n_{max}=N\log R/2|\log\xi|^2$.

\subsection{Miscellaneous Formula  for the Coherent States}

One can generate the coherent states $|j,z\ra$ for the representation of spin $j$ and $N=(2j-1)$ from the coherent states for the null-like representation of spin $j=\f12$ and $N=0$:
\be
|j,z\ra
\,=\,
\f1{(2L(z)^N\sqrt{N!}}(z^0a\dag-\bz^1 b)^N\,|\f12,z\ra\,.
\ee
One can check this by expanding the binomial and using the explicit definition of the coherent states in the basis labeled by the numbers of quanta. The interesting fact is that the operator $(z^0a\dag-\bz^1 b)$ behaves covariantly under the action of $\SU(1,1)$ as one can easily see from its commutation with the generators $\vJ$:
\be
U(z^0a\dag-\bz^1 b)U^{-1}
\,=\,
(U\vartriangleright z)^0a\dag-\overline{(U\vartriangleright z)^1} b\,.
\ee

\subsection{Spectrum and Eigenstates of the Boost Generator $K_y$}
\label{Ky}
\label{B7}

Let us look at eigenstates of the Lie algebra generators. The rotation generator $J_z$ gives the total energy of the two harmonic oscillators $a$ and $b$. It has a discrete positive spectrum and is diagonalized by the standard basis $|j,m\ra$ with $m\in j+\N$ defined above. On the other hand, the spectrum of a boost generator is purely continuous and is the entire real line.

Let us consider $K_y$, expressed in terms of $a$ and $b$. It admits a decoupled expression in terms of new oscillator operators:
\be
K_y=\f1{2i}\left(a\dag b\dag - ab\right)
=\f1{4i}\left(c\dag{}^2 -c^2 - d\dag{}^2 +d^2\right)
\ee
$$
a=\f1{\sqrt{2}}(c+d),
\quad
b=\f1{\sqrt{2}}(c-d),
\quad
[c,c\dag]=[d,d\dag]=1,
\quad
[c,d]=0\,.
$$
We can thus focus on the $c$-part of the operator and we define the Hermitian operator $\cD=(c\dag{}^2 -c^2)/2i$. This is the generator of Bogoliubov transformations on the oscillator $c$ and it maps coherent states to squeezed states:
\be
e^{-i\eta \cD} \,c\, e^{i\eta \cD}= (\cosh\eta \,c +\sinh\eta \,c\dag),
\quad
e^{-i\eta \cD} \,c\dag \,e^{i\eta \cD}= (\cosh\eta \,c\dag +\sinh\eta\, c)\,.
\ee
Using the standard quantization for the harmonic oscillator, we represent the creation and annihilation operators as functions acting on $\R$:
\be
c=\f1{\sqrt{2}}(x+\pp_x),\quad
c\dag=\f1{\sqrt{2}}(x-\pp_x)\,.
\ee
Then the operator $\cD$ turns out to be simply the dilatation operator acting on $x$:
\be
\cD=i\left[x\pp_x +\f12\right]\,.
\ee
Its spectrum is the real line is its eigenvectors are:
\be
\la x|\lambda\ra\,\equiv\,\f1{x^{\f12+i\lambda}},\quad
\cD\,|\lambda\ra\,=\,\lambda\,|\lambda\ra\,.
\ee

One can also consider the eigenvalue problem in the $|n_a,n_b\ra$ basis which we used to build the coherent states. For fixed $N=n_a-n_b$, let us act with $K_y$ on arbitrary states:
\be
K_y\sum_{n\in\N} (-1)^n\sqrt{\f{N!}{n!(n+N)!}} \alpha_n \,|n+N,n\ra
\,=\,
\f1{2i}\sum_{n\in\N} (-1)^n\sqrt{\f{N!}{n!(n+N)!}} (\alpha_{n+1}-n(n+N)\alpha_{n-1}) \,|n+N,n\ra\,.
\ee
Thus the coefficients of the eigenvector with eigenvalue $\lambda$ satisfy the following second order recursion relation:
\be
\forall n\ge0,\quad
\alpha_{n+1}(\lambda)=
2i\lambda\alpha_{n}(\lambda)+n(n+N)\alpha_{n-1}(\lambda)\,,
\ee
with initial conditions $\alpha_{-1}=0$ and arbitrary $\alpha_0$. Setting $\alpha_0=1$, the coefficients $\alpha_{n}(\lambda)$ will be polynomials of order $n$ in $2i\lambda$. It should be possible to map this recursion relation onto an orthogonal polynomial problem, but we do not investigate this direction further since we do not explicitly need the eigenvectors of $K_y$ but only the coherent states for the purpose of the work presented here. Nevertheless, the interested reader can refer to e.g. \cite{SU11} for more details on the representation theory of $\SU(1,1)$ and their recoupling.

\section{Classical FRW model with curvature and cosmological constant}
\label{AppC}

In this appendix we will review  the classical (unregularized) FRW model with intrinsic curvature $k=\pm 1$ and/or cosmological constant $\Lambda$, and coupled to a massless scalar field $\phi$. In the geometrodynamic variables $(a,\pi_a)$ the scalar constraint reads
\be
C=-\f{2\pi G}{3}\f{\pi_a^2}{a}-\f{3k}{8\pi G}a+\f{\Lambda}{8\pi G}a^3+\f{p_\phi^2}{2a^3}=0.
\ee

The canonical transformation between the above variables and the coefficients $c$ measuring the Ashtekar-Barbero connection and $p$ measuring the densitized triad in this general case is given by \footnote{In the following, for simplicity, we will assume that $p$ is positive, and therefore also $v$.}
\be
a=\sqrt{p}, \qquad \pi_a=-\frac{3}{4\pi G \gamma}\sqrt{p}\left(c-k\right)\,,
\ee
thus the constraint in these variables is given by \cite{AsS}
\begin{align}
C=\frac{1}{16\pi G}\left\{-\frac{6}{\gamma^2}\sqrt{p}\left[\left(c-k\right)^2+k\gamma^2\right]+2\Lambda p^{3/2}+8\pi G \f{p_\phi^2}{p^{3/2}}\right\}=0\,.
\end{align}
Introducing as before the canonical transformation to the (dimensionfull) variables $(v,b)$ given by
\be
p=\left(4\pi G v\right)^{2/3}\,, \qquad c=\gamma \left(4\pi G v\right)^{1/3} b\,,\\
\ee
we obtain
\be
C=-\f32 b^2v+\f{3k}{\gamma}\f{bv^{2/3}}{(4\pi G)^{1/3}}-\f3{8\pi G\gamma^2}(4\pi Gv)^{1/3}\left(k^2+k\gamma^2\right)+\f{\Lambda}{2}v+\f{p_\phi^2}{8\pi G v}=0.
\ee
Now we can deparameterize the system, solving the Hamiltonian constraint for the momentum of the field:
\be\label{const}
p_\phi^2=12\pi G \left[b^2v^2-\f{2k}{\gamma}\f{bv^{5/3}}{(4\pi G)^{1/3}}+\f{k^2+k\gamma^2}{\gamma^2(4\pi G)^{2/3}}v^{4/3}-\f{\Lambda}{3}v^2\right].
\ee
The square root of this expression give us the Hamiltonian of the system, that generates evolution in the internal time $\phi$. We note that the equation of motion of $\phi$ in terms of the proper time $t$ is given by $d{\phi}/dt=\{\phi,C\}$, so that the relation between the proper time and the internal time is
\be
t=\f{4\pi G}{p_\phi}\int v(\phi)d\phi.
\ee

Let us integrate the equations of motion for the simple
cases in which either the curvature or the cosmological constant vanish.

\subsection{Flat model with cosmological constant}
\label{C2}
The Hamiltonian particularizes to $p_\phi\equiv H^\pm=\pm \sqrt{12\pi G}v \sqrt{b^2-\f{\Lambda}{3}}$. To simplify the notation we introduce $\tau=\sqrt{12\pi G}\phi$. The resulting equations of motion are
\begin{align}
\partial_\tau v=\mp \f{bv}{\sqrt{b^2-\f{\Lambda}{3}}}\,,\qquad
\partial_\tau b=\pm\sqrt{b^2-\f{\Lambda}{3}}\,,
\end{align}
with a simple Friedmann equation:
\be
\pp^2_\tau v = v + \f{8\pi G\Lambda}{p_\phi^2} v^3.
\ee
We distinguish two kind of solutions depending on the sign of $\Lambda$:
\begin{itemize}
 \item $\Lambda> 0$: de Sitter Universe
\be
v(\tau)=\f{p_\phi}{\sqrt{4\pi G \Lambda}}\f1{|\sinh(\tau-\tau_o)|}\,,\qquad b(\tau)=\pm\sqrt{\f{\Lambda}{3}}\cosh(\tau-\tau_o)\,.
\ee
The matter density $\rho_\phi=p_\phi^2/(2V^2)$ reads
\be
\rho_\phi(\tau)=\f{\Lambda}{8\pi G}\sinh^2(\tau-\tau_o).
\ee
There are two branches of solutions (see fig. \ref{plot_Lpos}): for $\tau\in(-\infty,\tau_o)$ the solutions represent a universe that expands from a big bang singularity  till the matter density vanishes and the volume diverges;  for $\tau\in(\tau_o,\infty)$ the solutions represent a universe that contracts from infinite volume and vanishing density till a big crunch singularity.
From \eqref{propert} and the trajectory $v(\tau)$ we obtain that the proper time goes as
\be
t\sim\pm\ln\left|\tanh\f{\tau-\tau_o}{2}\right|,
\ee
where the positive sign corresponds to the branch $\tau\in(\tau_o,\infty)$ and the negative sign corresponds to the branch $\tau\in(-\infty,\tau_o)$. Moreover, $v(t)\propto|\sinh c t|$, with $c$ a constant.
Therefore in proper time we have a contracting branch for $t\in(-\infty,0)$ and a expanding branch for $t\in(0,\infty)$, and the instant $t=0$ leads to a curvature singularity.

\begin{figure}[h]
\begin{center}
\includegraphics[height=35mm]{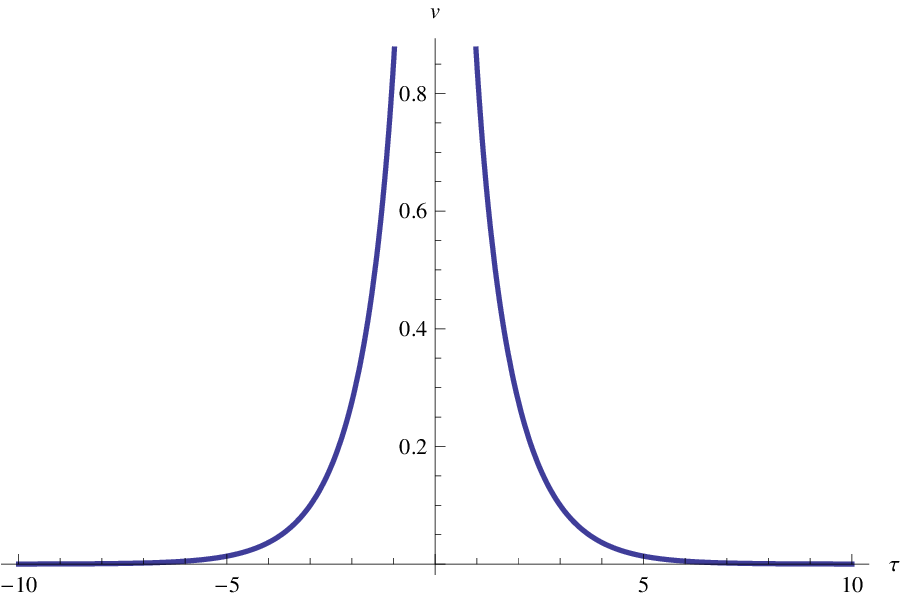}
\hspace{10mm}
\includegraphics[height=35mm]{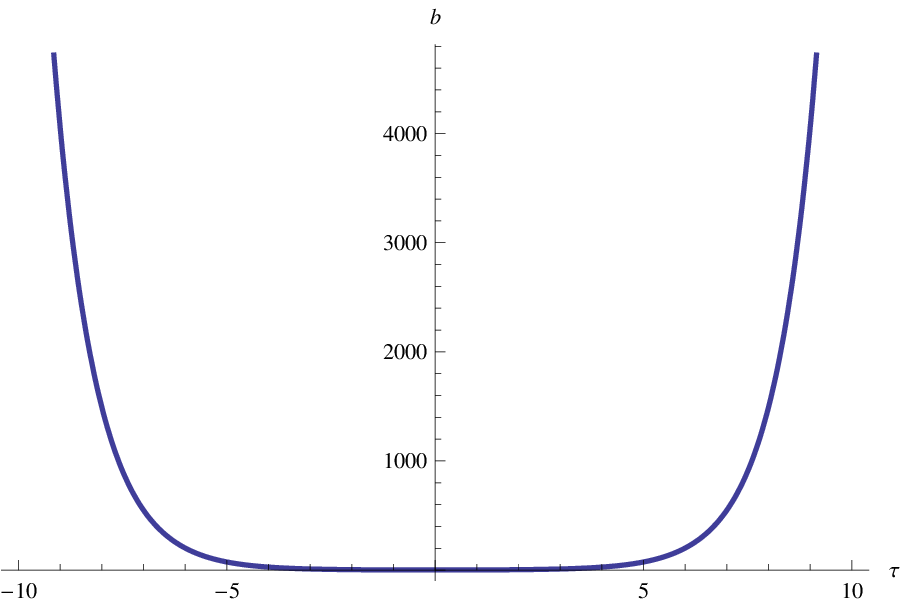}
\caption{Plots of the volume $v$ (on the left) and its conjugate momentum $b$ (on the right) evolving as functions of the internal time $\tau$, for $\tau_0=0$, for flat FRW cosmology with positive cosmological constant $\Lambda>0$. We have two branches: an expanding one starting with a big bang singularity and a contracting one ending with a big crunch.
\label{plot_Lpos}}
\end{center}
\end{figure}

A successful regularized dynamics should cure this singularity matching the two branches in a single one for the range $t\in(-\infty,\infty)$, representing a universe that contracts till a bouncing point at $t=0$ with positive volume and finite density, where it starts expanding. Such a bouncing behavior is achieved by the loop quantization. For this model the loop quantization has been thoroughly analyzed in \cite{AsP} (see also references therein), where the resulting classical effective dynamics is also reviewed.
\\

\item $\Lambda< 0$: anti-de Sitter Universe
\be
v(\tau)=\f{p_\phi}{\sqrt{4\pi G |\Lambda|}}\f1{\cosh(\tau-\tau_o)}\,,\qquad b(\tau)=\pm\sqrt{\f{|\Lambda|}{3}}\sinh(\tau-\tau_o)\,.
\ee
with matter density
\be
\rho_\phi(\tau)=\f{|\Lambda|}{8\pi G}\cosh^2(\tau-\tau_o).
\ee
These solutions represent a recollapsing universe (see fig. \ref{plot_Lneg}): it expands from a big bang singularity (at $\tau\rightarrow -\infty$) till the matter density reaches a minimum value equal to $|\Lambda|/8\pi G$ and the volume reaches a maximum value equal to $\sqrt{4\pi G/\Lambda}p_\phi$ (at $\tau=\tau_o$), moment at which the universe starts contracting till it reaches a big crunch singularity (at $\tau\rightarrow\infty$).

\begin{figure}[h]
\begin{center}
\includegraphics[height=35mm]{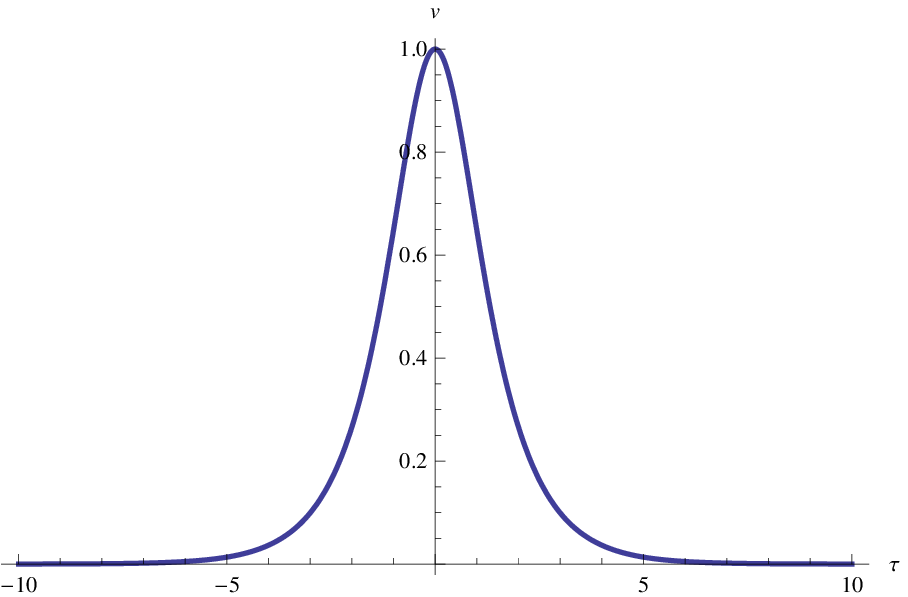}
\hspace{10mm}
\includegraphics[height=35mm]{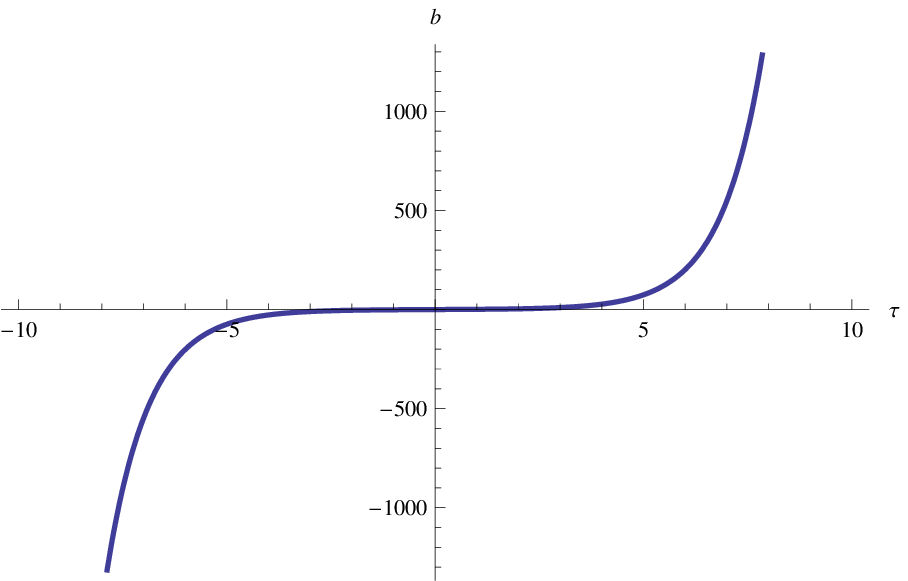}
\caption{Plots of the volume $v$ (on the left) and its conjugate momentum $b$ (on the right) evolving as functions of the internal time $\tau$, for $\tau_0=0$, for flat FRW cosmology with negative cosmological constant $\Lambda<0$. The universe starts in a big bang, expand to maximal volume and collapses again.
\label{plot_Lneg}}
\end{center}
\end{figure}

From \eqref{propert} and the trajectory $v(\tau)$ we obtain that the proper time goes as
\be
t\sim\arctan\left(e^{\tau-\tau_o}\right).
\ee
Moreover, $v(t)\propto\sin 2ct$, with $c$ a constant.
Therefore in proper time we also have a recollapsing universe, that starts at $t=0$ in a big bang singularity and dies for a finite value of the proper time in a big crunch.

A successful regularized dynamics should cure both big bang and big crunch singularities, smoothing them in terms of a bounce and making the evolution periodic. This behavior is again achieved by the loop quantization, which in this case has been thoroughly analyzed in \cite{BP} (see also references therein).
\end{itemize}

\subsection{Curved model without cosmological constant}
\label{C3}

In this case it is more convenient to solve the equations of motion for $a$ and $\pi_a$ and then to get the evolution for $v$ and $b$ employing the canonical transformation
\be
v=\f{a^3}{4\pi G}\,,\qquad b=-\f{4\pi G}{3}\f{\pi_a}{a^2}+\f{k}{\gamma a}.
\ee
We note that in this case the definition of $b$ involves an extra term, in comparison with the flat case, that depends on the Immirzi parameter $\gamma$. As a consequence, the trajectory for $b$ in time will depend explicitly on $\gamma$.
The Hamiltonian in $(a,\pi_a)$ variables reads
\be
p_\phi\equiv H^\pm=\pm\sqrt{\f{4\pi G}{3}\pi_a^2a^2+\f{3k}{4\pi G}a^4}.
\ee
The resulting equations of motion are
\be
\partial_\phi a=\f{4\pi G}{3p_\phi}\pi_a a^2\,,\qquad
\partial_\phi \pi_a=-\f{a}{p_\phi}\left(\f{4\pi G}{3}\pi_a^2 +\f{3k}{2\pi G}a^2\right)\,.
\ee
To get the solutions, we first use the conservation of the momentum $p_\phi$ to get the expression of $\pi_a$ in terms of $a$ and $p_\phi$, which allows us to solve the equation of motion for $a$, and consequently also the evolution for $\pi_a$. Considering only the positive branch for the scale factor $a$ the solutions are the following, depending on the sign of the curvature:

\begin{itemize}
\item $k=1$:
\begin{align}
a(\phi)&=\left(\f{4\pi G}{3}\right)^{1/4}\sqrt{\f{p_\phi}{\cosh\left[\sqrt{\f{16\pi G}{3}}(\phi-\phi_o)\right]}}\,,\\
\pi_a(\phi)&=-\left(\f{3}{4\pi G}\right)^{3/4}\sqrt{\f{p_\phi}{\cosh\left[\sqrt{\f{16\pi G}{3}}(\phi-\phi_o)\right]}}\sinh\left[\sqrt{\f{16\pi G}{3}}(\phi-\phi_o)\right].
\end{align}
Using $\tau\equiv \sqrt{12\pi G}\phi$ as in previous cases, for $v(\tau)$ and $b(\tau)$ we obtain:
\begin{align}
v(\tau)&=\f1{4\pi G}\left(\f{4\pi G}{3}\right)^{3/4}\left(\f{p_\phi}{\cosh\left[\f{2}{3}(\tau-\tau_o)\right]}\right)^{3/2}\,,\\
b(\tau)&=\left(\f{3}{4\pi G}\right)^{1/4}\sqrt{\f{\cosh\left[\f{2}{3}(\tau-\tau_o)\right]}{p_\phi}}\left\{\sinh\left[\f{2}{3}(\tau-\tau_o)\right]+\f1{\gamma}\right\}.
\end{align}
Therefore, the matter density is given by
\be
\rho_\phi(\tau)=\f{1}{2p_\phi}\left(\f{3}{4\pi G}\right)^{3/2}\cosh^3\left[\f{2}{3}(\tau-\tau_o)\right].
\ee
These solutions represent a universe that expands from a big bang singularity (at $\tau\rightarrow -\infty$) till the matter density reaches a minimum value equal to $\f{1}{2p_\phi}\left(\f{3}{4\pi G}\right)^{3/2}$ and the volume reaches a maximum value equal to $\left(\f{4\pi G}{3}\right)^{3/4}p_\phi^{3/2}$ (at $\tau=\tau_o$), moment at which the universe starts contracting till it reaches a big crunch singularity (at $\tau\rightarrow\infty$).
Both the loop quantization and the resulting effective dynamics of this model have been studied in \cite{apsv}, see also \cite{pol} for some technical aspects of the quantization.
\\

\item $k=-1$:
\begin{align}
a(\phi)&=\left(\f{4\pi G}{3}\right)^{1/4}\sqrt{\f{p_\phi}{\left|\sinh\left[\sqrt{\f{16\pi G}{3}}(\phi-\phi_o)\right]\right|}}\,,\\
\pi_a(\phi)&=-\left(\f{3}{4\pi G}\right)^{3/4}\sqrt{\f{p_\phi}{\left|\sinh\left[\sqrt{\f{16\pi G}{3}}(\phi-\phi_o)\right]\right|}}\cosh\left[\sqrt{\f{16\pi G}{3}}(\phi-\phi_o)\right].
\end{align}
Using $\tau\equiv \sqrt{12\pi G}\phi$ as in previous cases, for $v(\tau)$ and $b(\tau)$ we obtain:
\begin{align}
v(\tau)&=\f1{4\pi G}\left(\f{4\pi G}{3}\right)^{3/4}\left(\f{p_\phi}{\left|\sinh\left[\f{2}{3}(\tau-\tau_o)\right]\right|}\right)^{3/2}\,,\\
b(\tau)&=\left(\f{3}{4\pi G}\right)^{1/4}\sqrt{\f{\left|\sinh\left[\f{2}{3}(\tau-\tau_o)\right]\right|}{p_\phi}}\left\{\cosh\left[\f{2}{3}(\tau-\tau_o)\right]+\f1{\gamma}\right\}.
\end{align}
Therefore, the matter density is given by
\be
\rho_\phi(\tau)=\f{1}{2p_\phi}\left(\f{3}{4\pi G}\right)^{3/2}\left|\sinh\left[\f{2}{3}(\tau-\tau_o)\right]\right|^3.
\ee

As in the positive cosmological constant term, there are two branches of solutions: for $\tau\in(-\infty,\tau_o)$ the solutions represent a universe that expands from a big bang singularity  till the matter density vanishes and the volume diverges;  for $\tau\in(\tau_o,\infty)$ the solutions represent a universe that contracts from infinite volume and vanishing density till a big crunch singularity.
For the loop quantization of this model we refer the reader to \cite{v,s}.
\end{itemize}


\end{document}